%% file: quant_synergy.tex
\documentclass[12pt, final]{article}

\usepackage{fullpage}

\usepackage{virgil_env}	
\usepackage{fullpage}
\input{this_symbols.tex}  
\input{this_authors.tex}	

\usepackage{wasysym}

\title{Quantifying synergistic mutual information}

\begin{document}
\maketitle


\begin{abstract}
Quantifying cooperation or synergy among random variables in predicting a single target random variable is an important problem in many complex systems.  We review three prior information-theoretic measures of synergy and introduce a novel synergy measure defined as the difference between the whole and the union of its parts.  We apply all four measures against a suite of binary circuits to demonstrate that our measure alone quantifies the intuitive concept of synergy across all examples.  We show that for our measure of synergy that independent predictors can have positive redundant information.
\end{abstract}
 
\section{Introduction}
\label{sect:intro}
Synergy is a fundamental concept in complex systems that has received much attention in computational biology \cite{narayanan05, balduzzi-tononi-08}.  Several papers \cite{berry03, bell03, nirenberg01, plw-10} have proposed measures for quantifying synergy, but there remains no consensus which measure is most valid.

The concept of synergy spans many fields and theoretically could be applied to any non-subadditive function.  But within the confines of Shannon information theory, synergy---or more formally, \emph{synergistic information}---is a property of a set of $n$  random variables $\setX = \{X_1, X_2, \ldots, X_n\}$ cooperating to predict (reduce the uncertainty of) a single target random variable $Y$.

One clear application of synergistic information is in computational genetics.  It is well understood that most phenotypic traits are influenced not only by single genes but by interactions among genes---for example, human eye-color is cooperatively specified by more than a dozen genes\cite{white11}.  The magnitude of this ``cooperative specification'' is the synergistic information between the set of genes $\setX$ and a phenotypic trait $Y$.  Another application is neuronal firings where potentially thousands of presynaptic neurons influence the firing rate of a single post-synaptic (target) neuron.  Yet another application is discovering the ``informationally synergistic modules'' within a complex system.  

The prior literature\cite{schneidman-03,anas2007} has termed several distinct concepts as ``synergy''.  This paper defines synergy as how much the whole is greater than (the union of) its atomic elements.\footnote{The techniques here are unrelated to the information geometry prospective provided by \cite{amari99}.  The well-known ``total correlation'' measure\cite{han78}, does not satisfy the desired properties for a measure of synergy.}

The prior works on Partial Information Decomposition \cite{plw-10, polani12, bertschinger12, lizier13} start with properties that a measure of redundant information, $\Icap$ satisfies and builds a measure of synergy from $\Icap$.  Although this paper deals directly with measures of synergy on ``easy'' examples, we are immensely sympathetic to this approach.  Our proposed measure of synergy does give rise to an $\Icap$ measure.

\clearpage

The properties our $\Icup$ satisfies are discussed in Appendix \ref{appendix:axioms}.  For pedagogical purposes all examples are \emph{deterministic}, however, these methods equally apply to non-deterministic systems.

\subsection{Notation}
We use the following notation throughout.  Let

\begin{description}
	\item[\hspace{0.4in} $n$:] The number of predictors $X_1, X_2, \ldots, X_n$.  $n \geq 2$.
	\item[\hspace{0.4in} $\X$:] The \emph{joint} random variable (coalition) of all $n$ predictors $X_1 X_2 \ldots X_n$.	
	\item[\hspace{0.4in} $X_i$:] The $i$'th predictor random variable (r.v.). $1 \leq i \leq n$.
	\item[\hspace{0.4in} $\setX$:] The \emph{set} of all $n$ predictors $\left\{ X_1, X_2, \ldots, X_n \right\}$.
	\item[\hspace{0.4in} $Y$:] The \emph{target r.v.} to be predicted.
	\item[\hspace{0.4in} $y$:] A particular state of the target r.v. $Y$.
\end{description}

All random variables are discrete, all logarithms are $\log_2$, and all calculations are in \emph{bits}.  Entropy and mutual information are as defined by \cite{cover-thomas-91}, $\ent{X}~\equiv~\sum_{x \in X} \Prob{x} \log \frac{1}{\Pr(x)}$, as well as $\info{X}{Y}~\equiv~\sum_{x,y} \Prob{x,y} \log \frac{\Pr(x,y)}{\Pr(x)\Pr(y)}$.

\subsection{Understanding PI-diagrams}
\label{sect:pidiagrams}
Partial information diagrams (PI-diagrams), introduced by \cite{plw-10}, extend Venn diagrams to properly represent synergy.  Their framework has been invaluable to the evolution of our thinking on synergy.

A PI-diagram is composed of nonnegative \emph{partial information regions} (PI-regions).  Unlike the standard Venn entropy diagram in which the sum of all regions is the joint entropy $\ent{\X, Y}$, in PI-diagrams the sum of all regions (i.e. the space of the PI-diagram) is the mutual information $\info{\X}{Y}$.  PI-diagrams are immensely helpful in understanding how the mutual information $\info{\X}{Y}$ is distributed across the coalitions and singletons of $\setX$.\footnote{Formally, how the mutual information is distributed across the set of all nonempty antichains on the powerset of $\setX$\cite{mw-antichain,sperner74}.}

\begin{figure}[h!bt]
	\centering
	\subfloat[$n=2$]{ \includegraphics[width=1.75in]{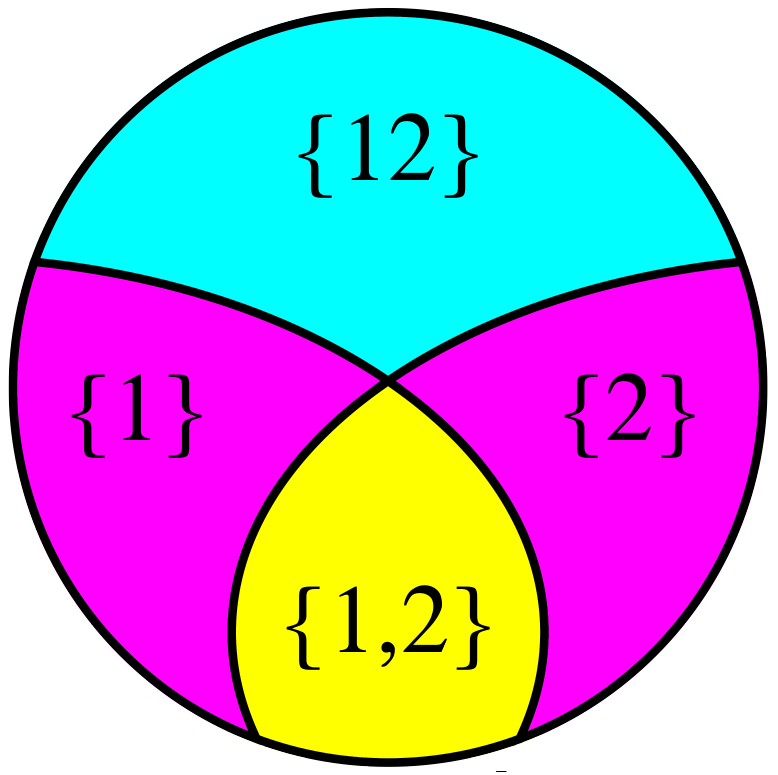} \label{fig:CYMa} }
	\subfloat[$n=3$]{ \includegraphics[width=3.45in]{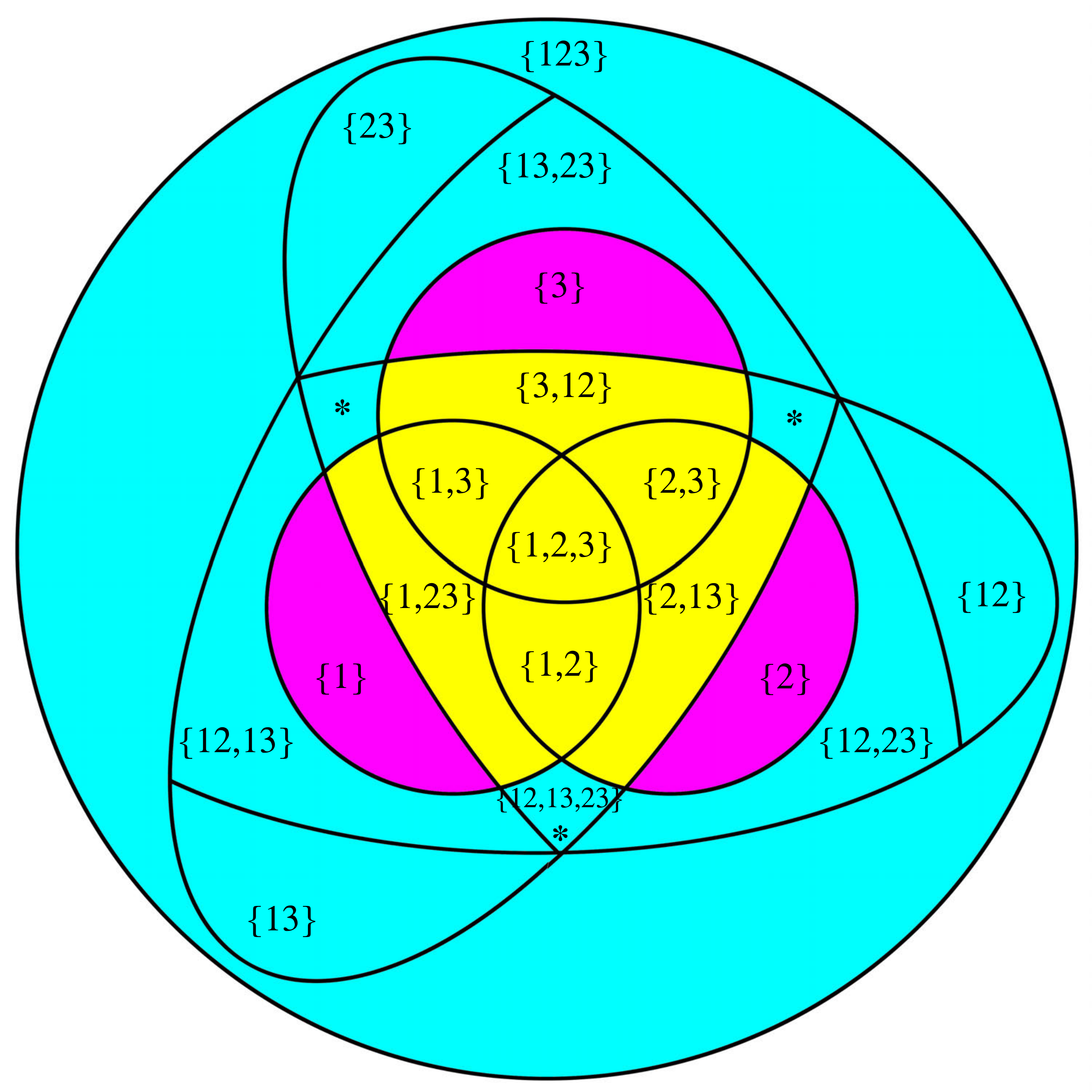} \label{fig:CYMb} }
	\caption{PI-diagrams for two and three predictors.  Each PI-region represents nonnegative information about $Y$.  A PI-region's color represents whether its information is redundant (yellow), unique (magenta), or synergistic (cyan).  To preserve symmetry, the PI-region ``$\{12,13,23\}$'' is displayed as three separate regions each marked with a ``*''.  All three *-regions should be treated as through they are a single region.}
	\label{fig:CYM}
\end{figure}

\begin{figure}[h!bt]
	\centering
	\subfloat[$\info{X_1}{Y}$]{ \includegraphics[width=0.86in]{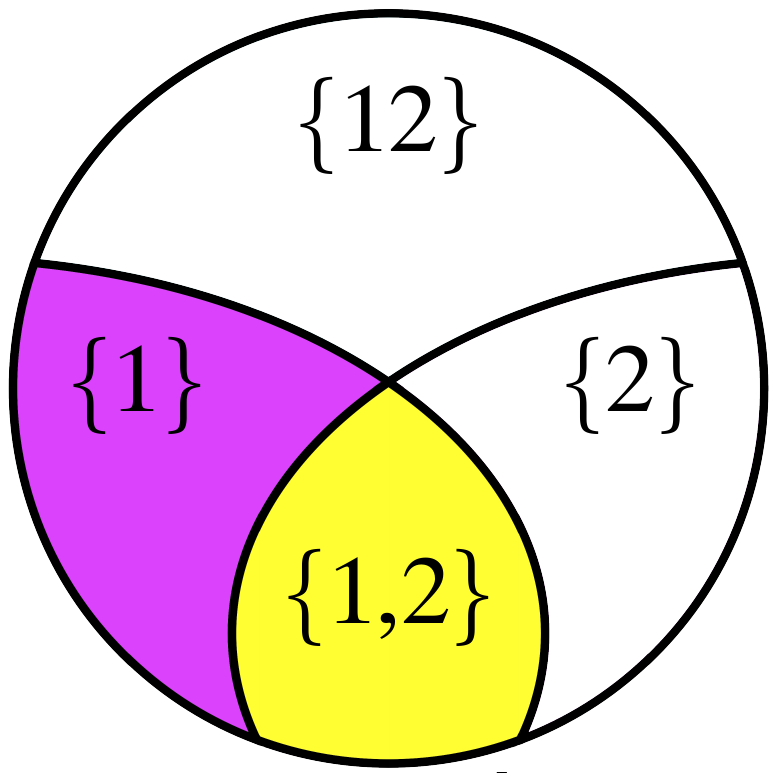} \label{fig:tutorial_a} }
	\subfloat[$\info{X_2}{Y}$]{ \includegraphics[width=0.86in]{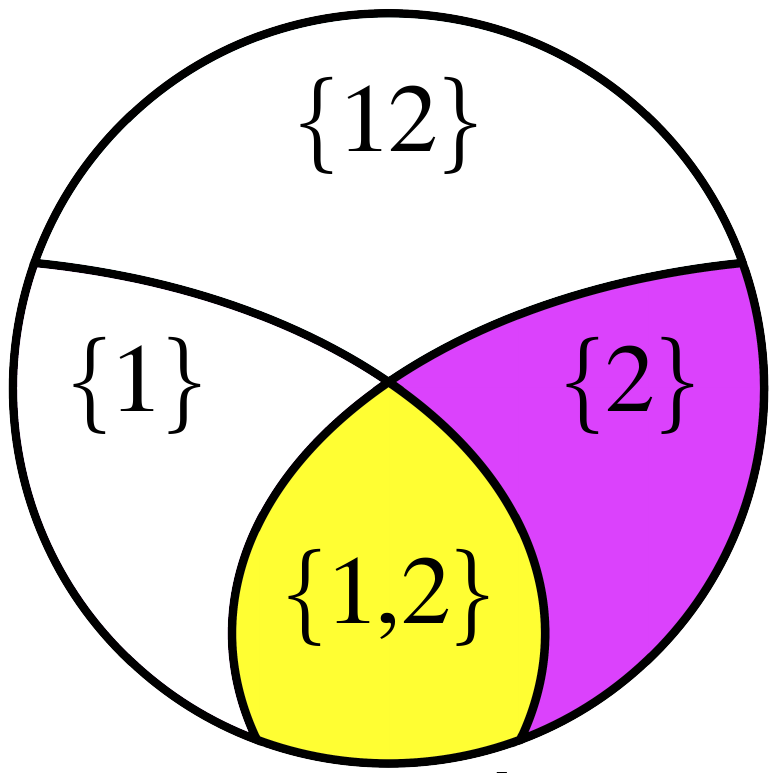} \label{fig:tutorial_b} }
	\subfloat[$\info{X_1}{Y|X_2}$]{ \includegraphics[width=0.86in]{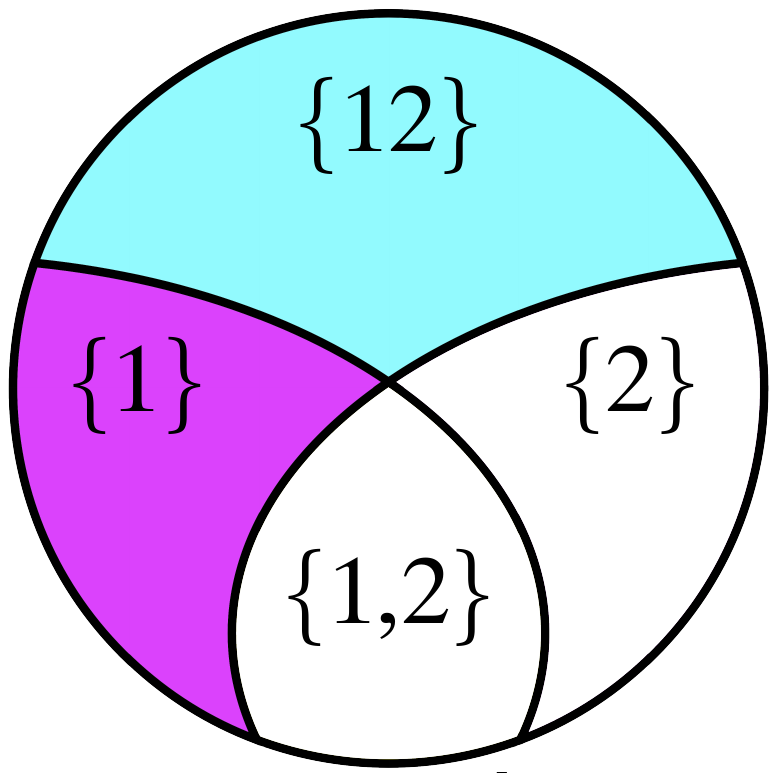} \label{fig:tutorial_c} }
	\subfloat[$\info{X_2}{Y|X_1}$]{ \includegraphics[width=0.86in]{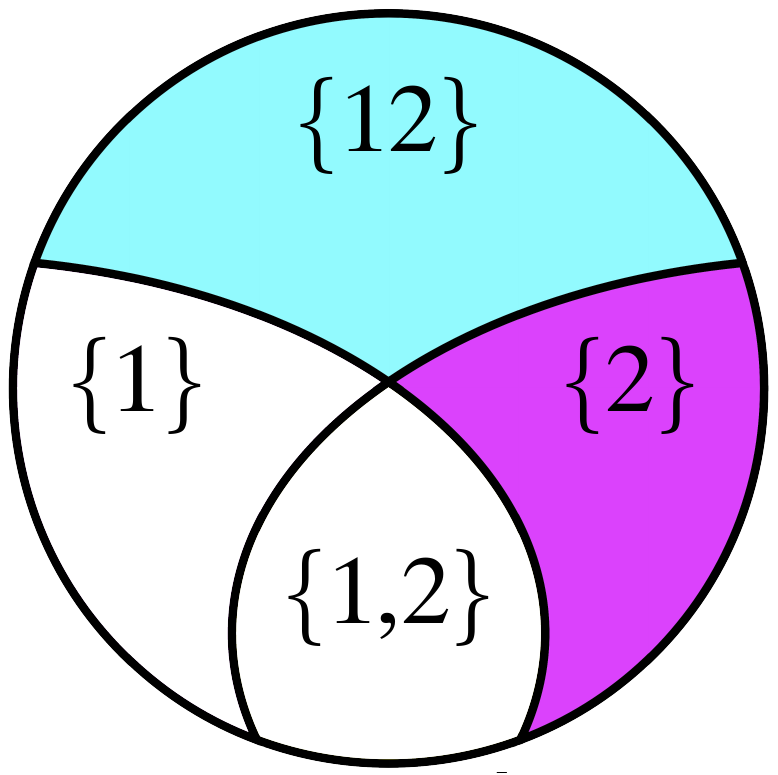} \label{fig:tutorial_d} }
	\subfloat[$\info{X_1 X_2}{Y}$]{ \includegraphics[width=0.86in]{PID2_CYM.pdf} \label{fig:tutorial_e} }
		\caption{PI-diagrams for $n=2$ representing standard informational quantities.}
	\label{fig:tutorial}
\end{figure}

\textbf{How to read PI-diagrams}. Each PI-region is uniquely identified by its ``set notation'' where each element is denoted solely by the predictors' indices.  For example, in the PI-diagram for $n=2$ (\figref{fig:CYMa}): $\{1\}$ is the information about $Y$ only $X_1$ carries (likewise \{2\} is the information only $X_2$ carries);  $\{1,2\}$ is the information about $Y$ that $X_1$ as well as $X_2$ carries, while $\{12\}$ is the information about $Y$ that is specified only by the coalition (joint random variable) $X_1 X_2$.  For getting used to this way of thinking, common informational quantities are represented by colored regions in \figref{fig:tutorial}.

The general structure of a PI-diagram becomes clearer after examining the PI-diagram for $n=3$ (\figref{fig:CYMb}).  All PI-regions from $n=2$ are again present.  Each predictor $\left( X_1, X_2, X_3 \right)$ can carry unique information (regions labeled \{1\}, \{2\}, \{3\}), carry information redundantly with another predictor (\{1,2\}, \{1,3\}, \{2,3\}), or specify information through a coalition with another predictor (\{12\}, \{13\}, \{23\}).  New in $n=3$ is information carried by all three predictors (\{1,2,3\}) as well as information specified through a three-way coalition (\{123\}).  Intriguingly, for three predictors, information can be provided by a coalition as well as a singleton (\{1,23\}, \{2,13\}, \{3,12\}) or specified by multiple coalitions (\{12,13\}, \{12,23\}, \{13,23\}, \{12,13,23\}).

\section{Information can be redundant, unique, or synergistic}
\label{sect:RUS}

Each PI-region represents an irreducible nonnegative slice of the mutual information $\info{\X}{Y}$ that is either:

\begin{enumerate}
	\item \textbf{Redundant}. Information carried by a singleton predictor as well as available somewhere else.  For $n=2$: \{1,2\}.  For $n=3$: \{1,2\}, \{1,3\}, \{2,3\}, \{1,2,3\}, \{1,23\}, \{2,13\}, \{3,12\}.
	\item \textbf{Unique}.  Information carried by exactly one singleton predictor and is available no where else.  For $n=2$: \{1\}, \{2\}.  For $n=3$: \{1\}, \{2\}, \{3\}.
	\item \textbf{Synergistic}. Any and all information in $\info{\X}{Y}$ that is not carried by a singleton predictor.  $n=2$: \{12\}.  For $n=3$: \{12\}, \{13\}, \{23\}, \{123\}, \{12,13\}, \{12,23\}, \{13,23\}, \{12,13,23\}.
\end{enumerate}

Although a single PI-region is either redundant, unique, or synergistic, a single state of the target can have any combination of positive PI-regions, i.e. a single state of the target can convey redundant, unique, and synergistic information.  This surprising fact is demonstrated in \figref{fig:RS}.

\subsection{Example Rdn: Redundant information}
\begin{figure}[h!b]
	\centering
	\begin{minipage}[c]{0.31\linewidth} \centering
		\subfloat[$\Prob{x_1,x_2,y}$]{ \begin{tabular}{ c | c c } \cmidrule(r){1-2}
		$X_1$ $X_2$ & $Y$ & \\
		\cmidrule(r){1-2} 
		\bin{r r} & \bin{r} & \quad \nicefrac{1}{2}\\
		\bin{R R} & \bin{R} & \quad \nicefrac{1}{2}\\
		\cmidrule(r){1-2} 
		\end{tabular} \label{fig:rdn_joint} }
		
	\end{minipage}
	\begin{minipage}[c]{0.31\linewidth} \centering	
		\subfloat[circuit diagram]{ \includegraphics[height=0.8in]{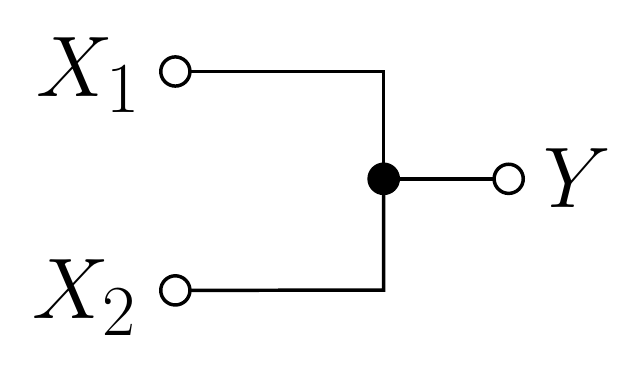} 	\label{fig:rdn_circuit} }

	\end{minipage}	
	\begin{minipage}[c]{0.31\linewidth} \centering	
	\subfloat[PI-diagram]{ \includegraphics[height=0.9in]{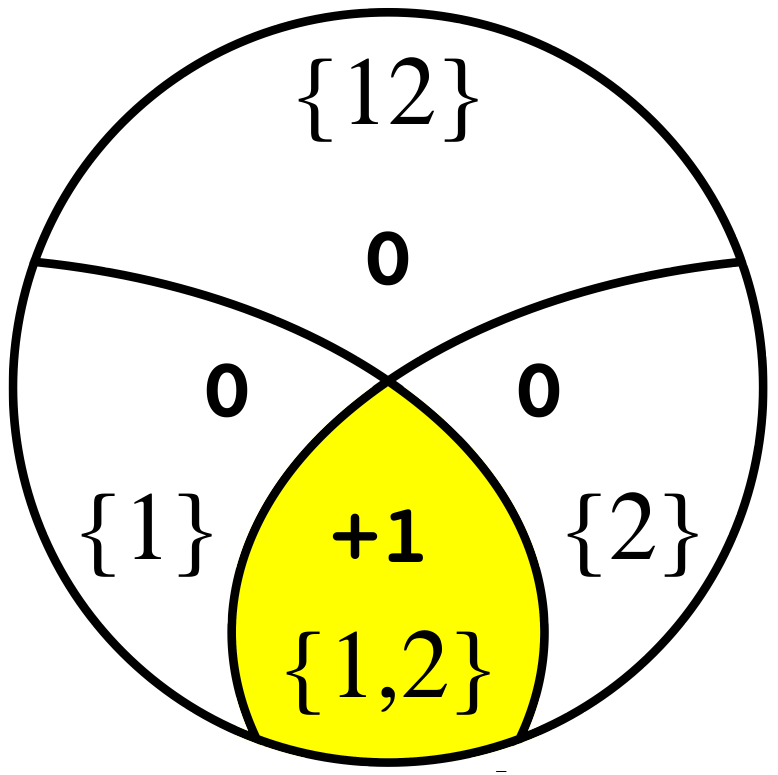} 	\label{fig:rdn_PID} }

	\end{minipage}
	\caption{Example \textsc{Rdn}.  \figref{fig:rdn_joint} shows the joint distribution of r.v.'s $X_1$, $X_2$, and $Y$, the joint probability $\Prob{x_1, x_2, y}$ is along the right-hand side of (a), revealing that all three terms are fully correlated.  \figref{fig:rdn_circuit} represents the joint distribution as an electrical circuit.  \figref{fig:rdn_PID} is the PI-diagram indicating that set $\{X_1, X_2\}$ has 1 bit of redundant information about $Y$. $\info{X_1 X_2}{Y}~=~\info{X_1}{Y}~=~\info{X_2}{Y}~=~\ent{Y}~=~1$ bit.}	
	\label{fig:exampleR}
\end{figure}

If $X_1$ and $X_2$ carry some identical\footnote{$X_1$ and $X_2$ providing identical information about $Y$ is different from providing the same \emph{magnitude} of information about $Y$, i.e. $\info{X_1}{Y} = \info{X_2}{Y}$.  Example \textsc{Unq} (\figref{fig:exampleU}) is an example where $\info{X_1}{Y} = \info{X_2}{Y} = 1$ bit yet $X_1$ and $X_2$ specify ``different bits'' of $Y$.  Providing the same magnitude of information about $Y$ is neither necessary or sufficient for providing some identical information about $Y$.} information  (reduce the same uncertainty) about $Y$, then we say the set $\setX = \{X_1, X_2\}$ has some \emph{redundant information} about $Y$.  \figref{fig:exampleR} illustrates a simple case of redundant information.  $Y$ has two equiprobable states: \bin{r} and \bin{R} (\bin{r}/\bin{R} for ``redundant bit'').  Examining $X_1$ or $X_2$ identically specifies one bit of $Y$, thus we say set $\setX = \{X_1, X_2\}$ has one bit of redundant information about $Y$.

\subsection{Example Unq: Unique information}
\begin{figure}[h!bt]
	\centering
	\begin{minipage}[c]{0.30\linewidth} \centering
	\subfloat[$\Prob{x_1,x_2,y}$]{\begin{tabular}{ c | c c} \cmidrule(r){1-2}
	$X_1$ $X_2$ &$Y$ \\
	\cmidrule(r){1-2} 
	\bin{a b} & \bin{ab} & \quad \nicefrac{1}{4}\\
	\bin{a B} & \bin{aB} & \quad \nicefrac{1}{4}\\
	\bin{A b} & \bin{Ab} & \quad \nicefrac{1}{4}\\
	\bin{A B} & \bin{AB} & \quad \nicefrac{1}{4}\\
	\cmidrule(r){1-2} 
	\end{tabular}	
	} \end{minipage}
	\begin{minipage}[c]{0.33\linewidth} \centering
	\subfloat[circuit diagram]{ \includegraphics[height=1.0in]{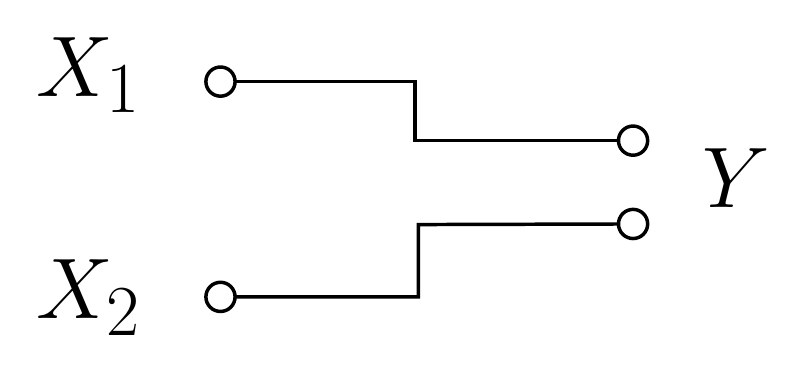} \label{fig:exampleUb} }
	\end{minipage}		
	\begin{minipage}[c]{0.33\linewidth} \centering
		\subfloat[PI-diagram]{ \includegraphics[height=1.0in]{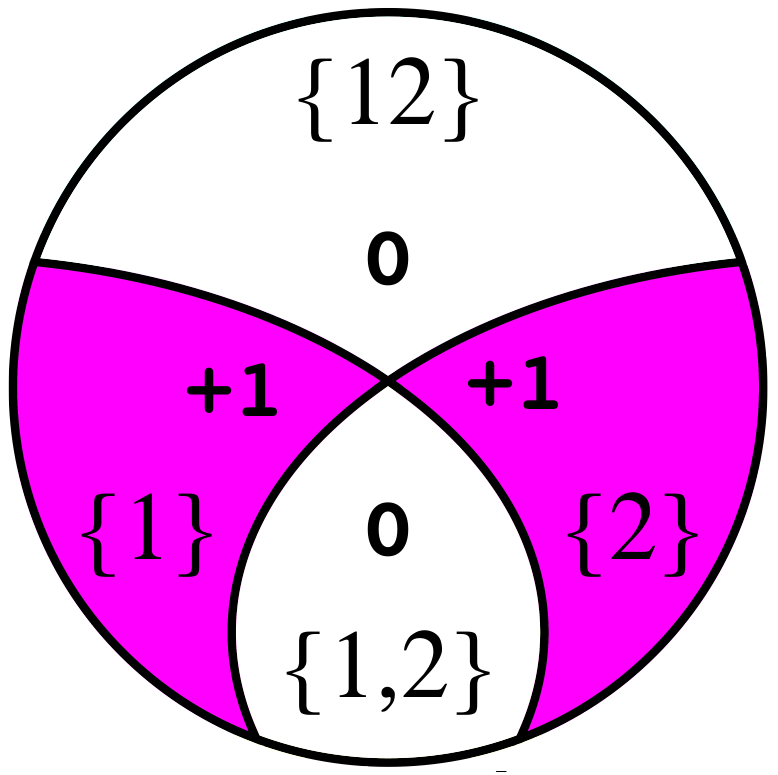} }
	\end{minipage}
	
	\caption{Example \textsc{Unq}.  $X_1$ and $X_2$ each uniquely specify a single bit of $Y$.  $\info{X_1 X_2}{Y}~=~\ent{Y}~=~2$ bits. The joint probability $\Prob{x_1, x_2, y}$ is along the right-hand side of (a).}
	\label{fig:exampleU}
\end{figure}

Predictor $X_i$ carries \emph{unique information} about $Y$ if and only if $X_i$ specifies information about $Y$ that is not specified by anything else (a singleton or coalition of the other $n-1$ predictors).  \figref{fig:exampleU} illustrates a simple case of unique information.  $Y$ has four equiprobable states: \bin{ab}, \bin{aB}, \bin{Ab}, and \bin{AB}.  $X_1$ uniquely specifies bit \bin{a}/\bin{A}, and $X_2$ uniquely specifies bit \bin{b}/\bin{B}.  If we had instead labeled the $Y\textnormal{-states}$: \bin{0}, \bin{1}, \bin{2}, and \bin{3}, $X_1$ and $X_2$ would still have strictly unique information about $Y$.  The state of $X_1$ would specify between $\{\bin{0}, \bin{1} \}$ and $\{ \bin{2}, \bin{3} \}$, and the state of $X_2$ would specify between $\{ \bin{0}, \bin{2} \}$ and $\{ \bin{1}, \bin{3} \}$---together fully specifying the state of $Y$.  Accepting the property \textbf{(Id)} from \cite{polani12} is sufficient but not necessary for the desired decomposition of example \textsc{Unq}.

\subsection{Example Xor: Synergistic information}
\begin{figure}[h!bt]
	\centering
	\begin{minipage}[c]{0.26\linewidth} \centering \subfloat[$\Prob{x_1,x_2,y}$]{ \begin{tabular}{ c | c c} \cmidrule(r){1-2}
	$X_1$ $X_2$ &$Y$ \\
	\cmidrule(r){1-2} 
	\bin{0 0} & \bin{0} & \quad \nicefrac{1}{4}\\
	\bin{0 1} & \bin{1} & \quad \nicefrac{1}{4}\\
	\bin{1 0} & \bin{1} & \quad \nicefrac{1}{4}\\
	\bin{1 1} &  \bin{0} & \quad \nicefrac{1}{4}\\
	\cmidrule(r){1-2} 
	\end{tabular} }
	\end{minipage}
	\begin{minipage}[c]{0.365\linewidth} \centering	
	\subfloat[circuit diagram]{ \includegraphics[height=0.65in]{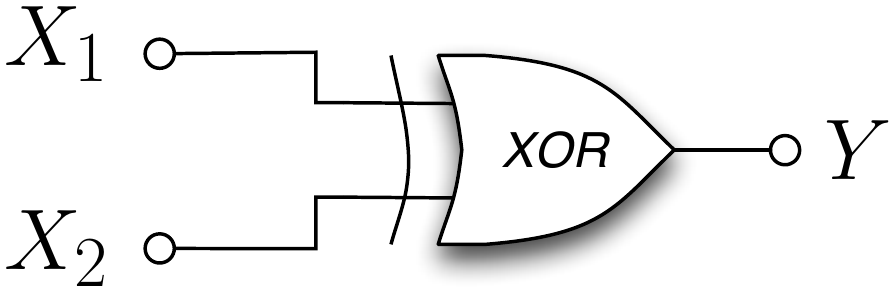} }
	\end{minipage}		
	\begin{minipage}[c]{0.33\linewidth} \centering
		\subfloat[PI-diagram]{ \includegraphics[height=1.0in]{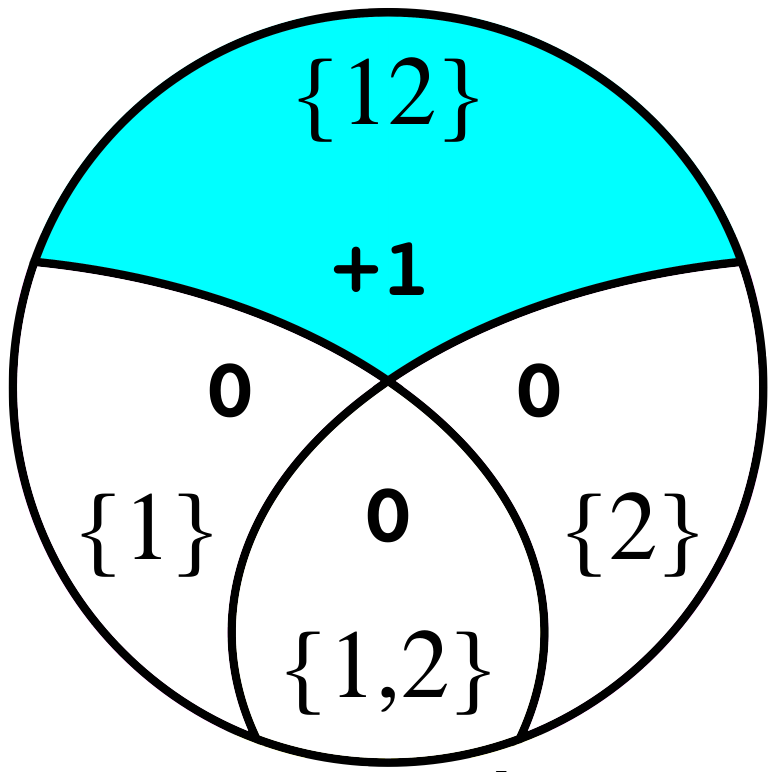} }
	\end{minipage}
	\caption{Example \textsc{Xor}.  $X_1$ and $X_2$ synergistically specify $Y$.  $\info{X_1 X_2}{Y}~=~\ent{Y}~=~1$ bit.  The joint probability $\Prob{x_1, x_2, y}$ is along the right-hand side of (a).}
	\label{fig:xor}
\end{figure}

A set of predictors $\setX=\{X_1, \ldots, X_n\}$ has synergistic information about $Y$ if and only if the whole ($\X$) specifies information about $Y$ that is not specified by any singleton predictor.  The canonical example of synergistic information is the \textsc{Xor}-gate (\figref{fig:xor}).  In this example, the whole $X_1 X_2$ fully specifies $Y$,
\begin{equation}
	\info{X_1 X_2}{Y} = \ent{Y} = 1 \textnormal{ bit,}
\end{equation}
but the singletons $X_1$ and $X_2$ specify \emph{nothing} about $Y$,
\begin{equation}
	\info{X_1}{Y} = \info{X_2}{Y} = 0 \textnormal{ bits.}
\end{equation}
With both $X_1$ and $X_2$ themselves having zero information about $Y$, we know that there can not be any redundant or unique information about $Y$---that the three PI-regions $\{1\}~=~\{2\}~=~\{1,2\}~=~0$ bits.  As the information between $X_1 X_2$ and $Y$ must come from somewhere, by elimination we conclude that $X_1$ and $X_2$ synergistically specify $Y$.

\section{Two examples elucidating properties of synergy}
To help the reader develop intuition for a proper measure of synergy we illustrate two desired properties of synergistic information with pedagogical examples derived from \textsc{Xor}.  Readers solely interested in the contrast with prior measures can skip to Section \ref{sect:examples}.

\subsection{Duplicating a predictor does not change synergistic information}
\begin{figure}[h!bt]
	\centering
	\begin{minipage}[c]{0.4\linewidth} \centering
\subfloat[$\Prob{x_1,x_2,x_3,y}$]{ \begin{tabular}{ c | c c } \cmidrule(r){1-2}
$\ \;X_1 \ X_2 \ X_3$  &$Y$ \\
\cmidrule(r){1-2} 
\bin{0 0 0} & \bin{0} & \quad \nicefrac{1}{4}\\
\bin{0 1 0} & \bin{1} & \quad \nicefrac{1}{4}\\
\bin{1 0 1} & \bin{1} & \quad \nicefrac{1}{4}\\
\bin{1 1 1} & \bin{0} & \quad \nicefrac{1}{4}\\
\cmidrule(r){1-2}
\end{tabular} }
\end{minipage} \begin{minipage}[c]{0.55\linewidth} \centering
	\subfloat[circuit diagram]{ \includegraphics[width=2.15in]{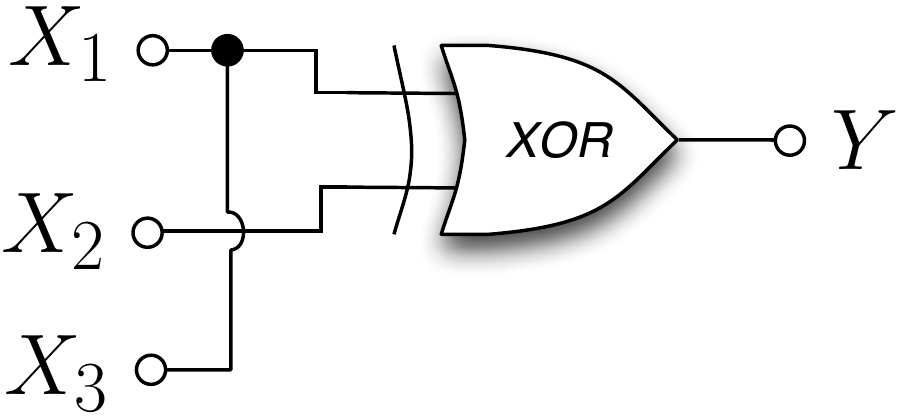} }
	\end{minipage}
\begin{minipage}[c]{\linewidth} \centering
	\subfloat[PI-diagram]{ \includegraphics[height=2.9in]{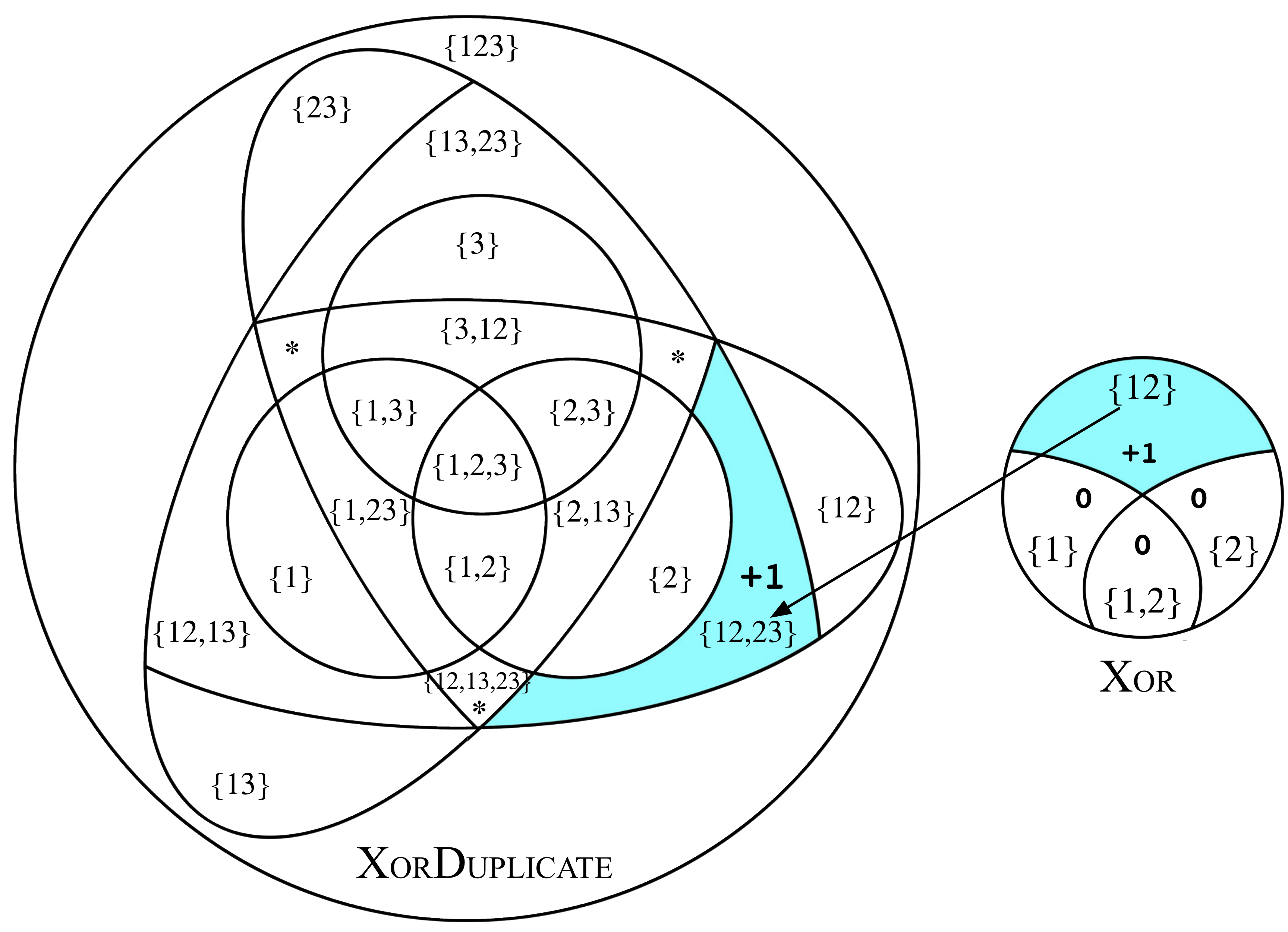} }
	\end{minipage}
	
	\caption{Example \textsc{XorDuplicate} shows that duplicating predictor $X_1$ as $X_3$ turns the single-coalition synergy $\{12\}$ into the multi-coalition synergy $\{12,23\}$.  After duplicating $X_1$, the coalition $X_3 X_2$ as well as coalition $X_1 X_2$ specifies $Y$.  Synergistic information is unchanged from \textsc{Xor}, $\info{X_3 X_2}{Y}~=~\info{X_1 X_2}{Y}~=~\ent{Y}~=~1$ bit.}
	\label{fig:XorDup}
\end{figure}

Example \textsc{XorDuplicate} (\figref{fig:XorDup}) adds a third predictor, $X_3$, a copy of predictor $X_1$, to \textsc{Xor}.  Whereas in \textsc{Xor} the target $Y$ is specified only by coalition $X_1 X_2$, duplicating predictor $X_1$ as $X_3$ makes the target equally specifiable by coalition $X_3 X_2$.

Although now two different coalitions identically specify $Y$, mutual information is invariant to duplicates, e.g. $\info{X_1 X_2 X_3 }{Y} = \info{X_1 X_2}{Y}$ bit.  Likewise for synergistic information to be likewise bounded between zero and the total mutual information $\info{\X}{Y}$, synergistic information must similarly be invariant to duplicates, e.g. the synergistic information between set $\{X_1, X_2\}$ and $Y$ must be the same as the synergistic information between $\{X_1, X_2, X_3\}$ and $Y$.  This makes sense because if synergistic information is defined as the information in the whole beyond its parts, duplicating a part does not increase the net information provided by the parts.  Altogether, we assert that \emph{duplicating a predictor does not change the synergistic information}.  Synergistic information being invariant to duplicated predictors follows from the equality condition of the monotonicity property $\mathbf{(M)}$ from \cite{bertschinger12}.\footnote{For a proof see Appendix \ref{appendix:proofs}.}

\subsection{Adding a new predictor can decrease synergy}
\begin{figure}[h!bt]
	\centering
	\begin{minipage}[c]{0.4\linewidth} \centering	
	\subfloat[$\Prob{x_1, x_2, x_3, y}$]{ \begin{tabular}{ c | c c } \cmidrule(r){1-2}
$\ \, X_1 \, X_2 \, X_3$  &$Y$ \\
\cmidrule(r){1-2} 
\bin{0 0 0} & \bin{0} & \quad \nicefrac{1}{4}\\
\bin{0 1 1} & \bin{1} & \quad \nicefrac{1}{4}\\
\bin{1 0 1} & \bin{1} & \quad \nicefrac{1}{4}\\
\bin{1 1 0} & \bin{0} & \quad \nicefrac{1}{4}\\
\cmidrule(r){1-2} 
\end{tabular} }
\end{minipage} \begin{minipage}[c]{0.55\linewidth} \centering
	\subfloat[circuit diagram]{ 	\includegraphics[width=1.6in]{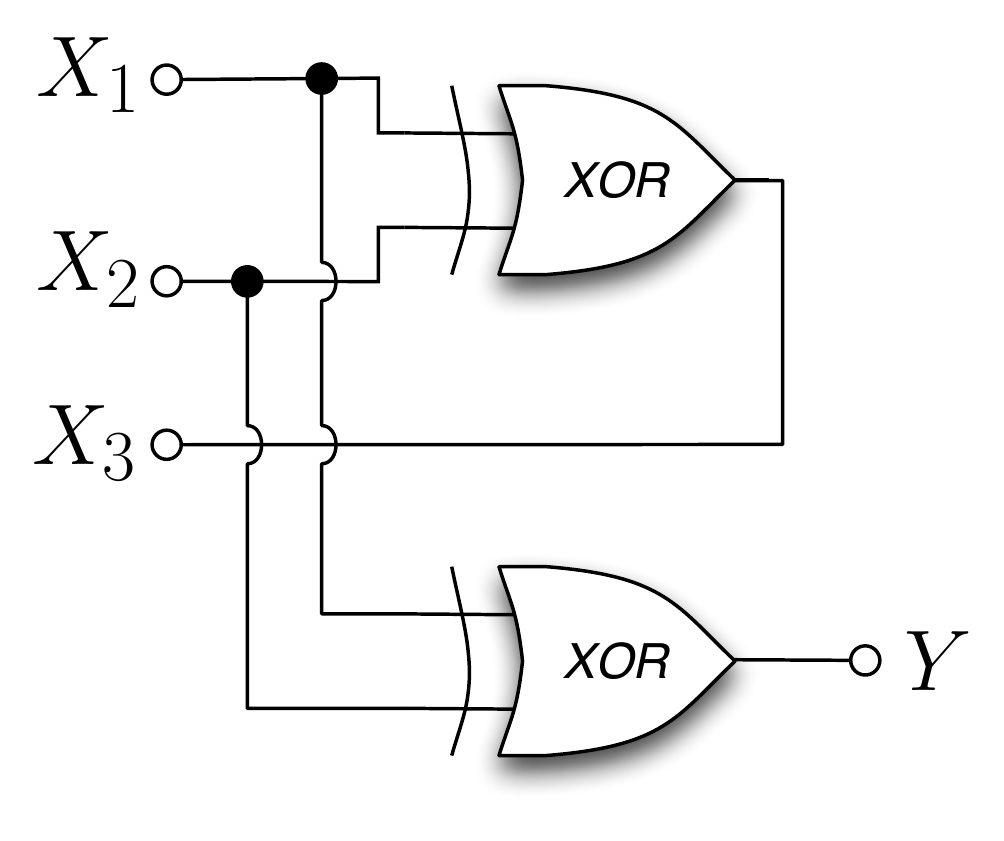} }
	\end{minipage}
\begin{minipage}[c]{\linewidth} \centering
	\subfloat[PI-diagram]{ 	\includegraphics[height=2.7in]{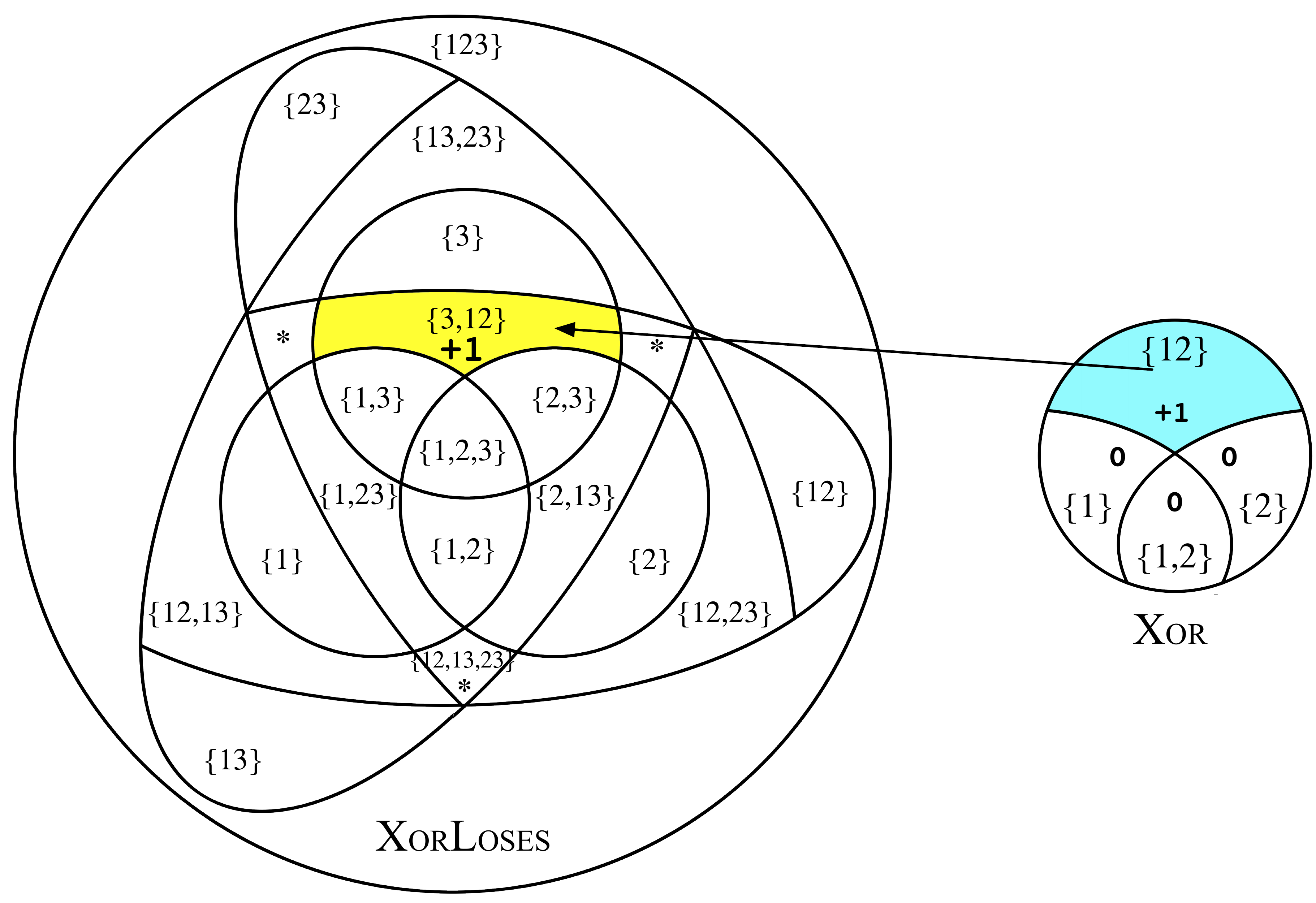} }
	\end{minipage}
	\caption{Example \textsc{XorLoses}.  Target $Y$ is fully specified by the coalition $X_1 X_2$ as well as by the singleton $X_3$.  $\info{X_1 X_2}{Y}~=~\info{X_3}{Y}~=~\ent{Y}~=~1$ bit.  Therefore the information synergistically specified by coalition $X_1 X_2$ is a redundant synergy.}
	\label{fig:xorallunique}
\end{figure}

Example \textsc{XorLoses} (\figref{fig:xorallunique}) adds a third predictor, $X_3$, to \textsc{Xor} and concretizes the distinction between synergy and ``redundant synergy''.  In \textsc{XorLoses} the target $Y$ has one bit of uncertainty and just as in example \textsc{Xor} the coalition $X_1 X_2$ fully specifies the target, $\info{X_1 X_2}{Y} = \ent{Y} = 1$ bit.  However, \textsc{XorLoses} has \emph{zero} intuitive synergy because the newly added singleton predictor, $X_3$, fully specifies $Y$ by itself.  This makes the synergy between $X_1$ and $X_2$ \emph{completely redundant}---everything the coalition $X_1 X_2$ specifies is now already specified by the singleton $X_3$.

\section{Prior measures of synergy}
\label{sect:examples}

\subsection{$\operatorname{I}_{\max}$ synergy: $\S_{\max}\left( \setX : Y \right)$}
$\opname{I}_{\max}$ synergy, denoted $\mathcal{S}_{\opname{max}}$, derives from \cite{plw-10}.  $\S_{\max}$ defines synergy as the whole beyond the state-dependent \emph{maximum} of its parts,

\begin{eqnarray}
	\mathcal{S}_{\opname{max}}\left( \setX : Y \right) &\equiv& \info{\X}{Y} - \opI_{\max}\left( \{X_1, \ldots, X_n\} : Y \right) \\
	 &=& \info{\X}{Y} - \sum_{y \in Y} \Prob{Y = y} \max_i \info{ X_i }{Y=y} \; ,
	\label{eq:Smax}
\end{eqnarray}
where $\info{X_i}{Y=y}$ is \cite{meister-99}'s ``specific-surprise'',
\begin{eqnarray}
	\info{X_i}{Y=y} &\equiv& \DKL{ \Prob{X_i|y} }{ \Prob{X_i} } \\
	&=& \sum_{x_i \in X_i} \Prob{x_i | y } \log \frac{ \Prob{x_i, y} }{ \Prob{x_i} \Prob{y} } \; .
\end{eqnarray}

There are two major advantages of $\S_{\max}$ synergy.  First, $\S_{\max}$ obeys the bounds of $0~\leq~\S_{\max}( \X~:~Y ) \leq \info{\X}{Y}$. Second, $\S_{\max}$ is invariant to duplicate predictors.  Despite these desired properties, $\S_{\max}$ sometimes miscategorizes merely unique information as synergistic.  This can be seen in example \textsc{Unq} (\figref{fig:exampleU}).  In example \textsc{Unq} the wires in \figref{fig:exampleUb} don't even touch, yet $\S_{\max}$ asserts there is one bit of synergy and one bit of redundancy---this is palpably strange.

A more abstract way to understand why $\mathcal{S}_{\max}$ overestimates synergy is to imagine a hypothetical example where there are exactly two bits of unique information for every state $y \in Y$ and no synergy or redundancy.  $\S_{\max}$ would be the whole (both unique bits) minus the \emph{maximum} over both predictors---which would be the $\max \left[ 1, 1\right] = 1$ bit.  The $S_{\max}$ synergy would then be $2 - 1 = 1$ bit of synergy---even though by definition there was no synergy, but merely two bits of unique information.

Altogether, we conclude that $\mathcal{S}_{\max}$ \emph{overestimates} the intuitive synergy by miscategorizing merely unique information as synergistic whenever two or more predictors have unique information about the target.

\subsection{WholeMinusSum synergy: $\WMS\left( \setX : Y \right)$}
\label{sect:WholeMinusSum}
The earliest known sightings of bivarate WholeMinusSum synergy (WMS) is \cite{richmond-93,Gat99} with the general case in \cite{chechik01}. WholeMinusSum synergy is a signed measure where a positive value signifies synergy and a negative value signifies redundancy.  WholeMinusSum synergy is defined by eq.~\eqref{eq:chechik1} and interestingly reduces to eq.~\eqref{eq:chechik3}---the difference of two \emph{total correlations}.\footnote{$\opname{TC}( X_1 ; \cdots ; X_n ) = - \ent{\X} + \sum_{i=1}^n \ent{X_i}$ per \cite{han78}.}

\begin{eqnarray}
\label{eq:chechik1}	\WMS\left( \setX : Y \right) &\equiv& \info{\X}{Y} - \sum_{i=1}^n \info{ X_i }{ Y } \\
\label{eq:chechik2}	&=& \sum_{i=1}^n \ent{X_i|Y} - \ent{\X|Y} - \left[ \sum_{i=1}^n \ent{X_i} - \ent{\X} \right] \\
\label{eq:chechik3}	&=& \opname{TC} \left( X_1 ; \cdots ; X_n \middle| Y \right) - \opname{TC}\left( X_1 ; \cdots ; X_n \right)
\end{eqnarray}

\begin{figure}[h!bt]
\centering
	\subfloat[$\WMS\left( \{ X_1, X_2\} : Y \right)$]{ \includegraphics[width=1.55in]{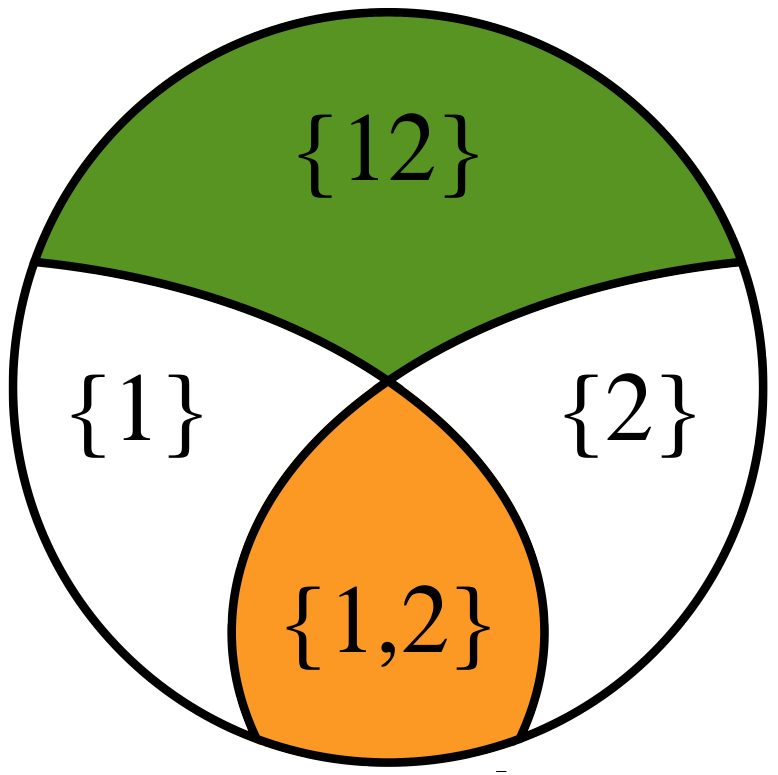} \label{fig:PID2_chechik} }
	\subfloat[$\WMS\left( \{ X_1, X_2, X_3\} : Y \right)$]{ \includegraphics[height=3.3in]{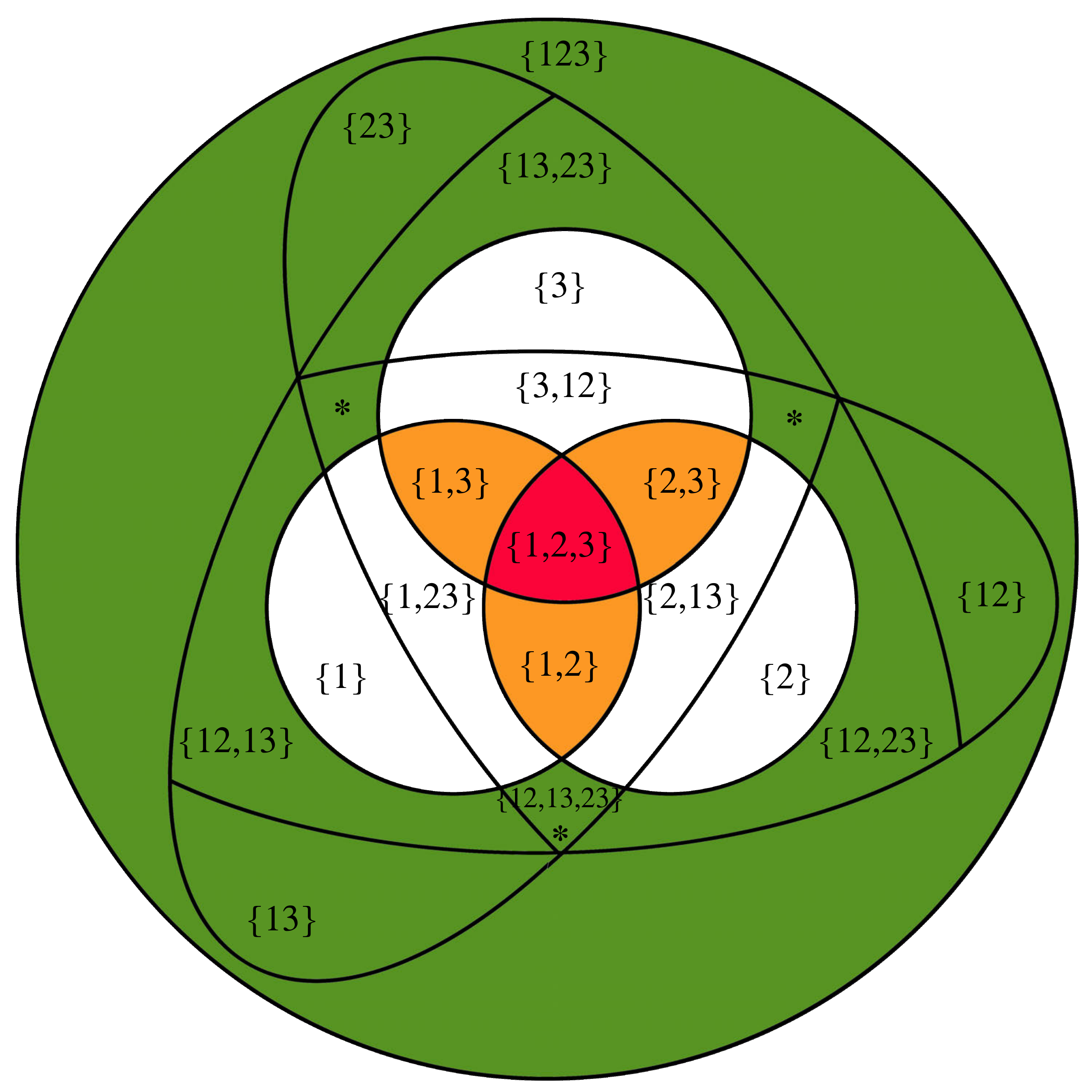} 
\label{fig:PID3_chechik} }
\caption{PI-diagrams illustrating WholeMinusSum synergy for $n=2$ (left) and $n=3$ (right).  For this diagram the colors denote the added and subtracted PI-regions.  $\WMS\left( \setX : Y \right)$ is the green PI-region(s), minus the orange PI-region(s), minus two times any red PI-region.}
\label{fig:chechik}
\end{figure}

\begin{figure}[h!bt]
	\centering
	\begin{minipage}[c]{0.28\linewidth} \centering \subfloat[$\Prob{x_1,x_2,y}$]{ \begin{tabular}{ c | c c } \cmidrule(r){1-2}
	$X_1$ $X_2$ & $Y$ \\
	\cmidrule(r){1-2} 
	\bin{r0 r0} & \bin{r0} & \quad \nicefrac{1}{8}\\
	\bin{r0 r1} & \bin{r1} & \quad \nicefrac{1}{8}\\
	\bin{r1 r0} & \bin{r1} & \quad \nicefrac{1}{8}\\
	\bin{r1 r1} & \bin{r0} & \quad \nicefrac{1}{8}\\
	\addlinespace
	\bin{R0 R0} & \bin{R0} & \quad \nicefrac{1}{8}\\
	\bin{R0 R1} & \bin{R1} & \quad \nicefrac{1}{8}\\
	\bin{R1 R0} & \bin{R1} & \quad \nicefrac{1}{8}\\
	\bin{R1 R1} & \bin{R0} & \quad \nicefrac{1}{8}\\
	\cmidrule(r){1-2} 
	\end{tabular} }
	\end{minipage}
	\begin{minipage}[c]{0.37\linewidth} \centering	
	\subfloat[circuit diagram]{ \includegraphics[width=1.8in]{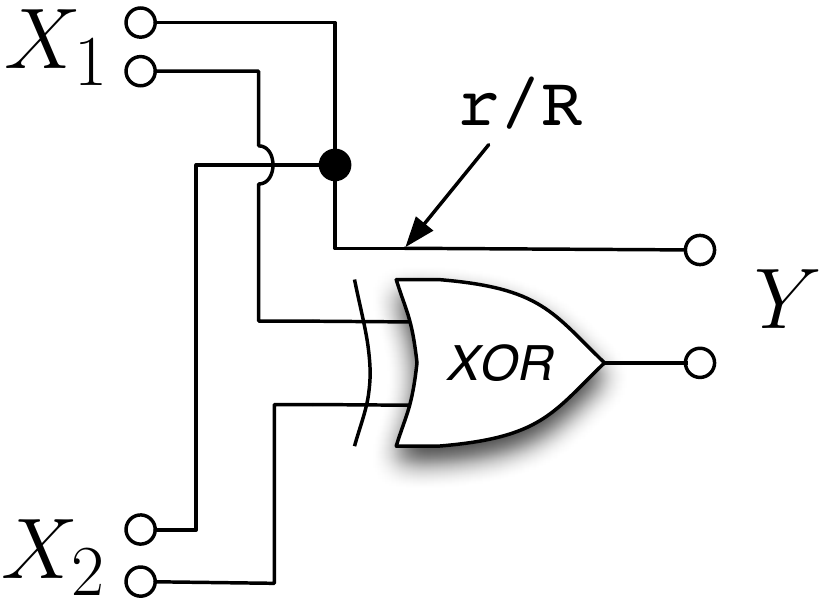} }
	\end{minipage}		
	\begin{minipage}[c]{0.32\linewidth} \centering
		\subfloat[PI-diagram]{ \includegraphics[width=1.7in]{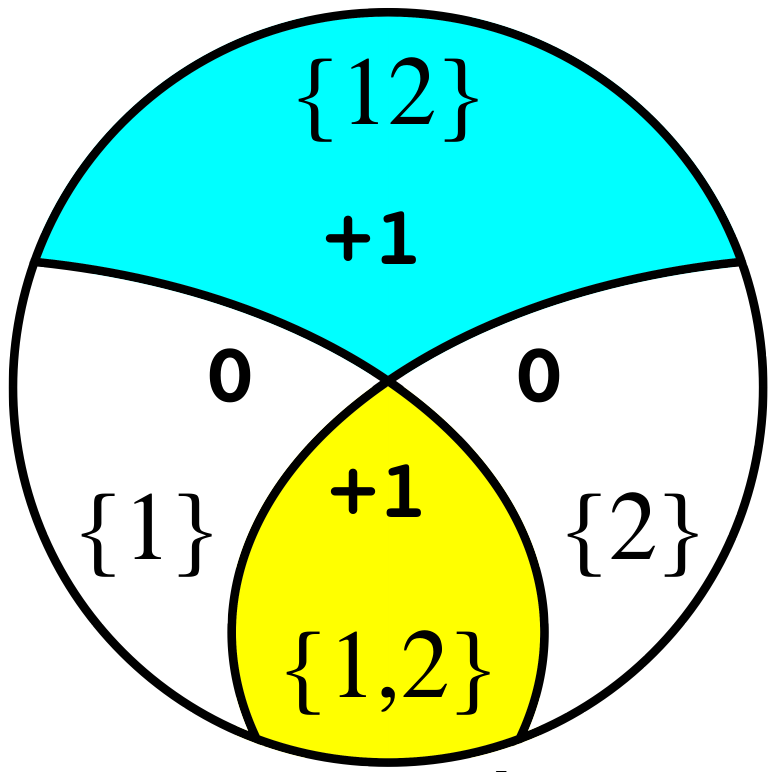} }
	\end{minipage}
	\caption{Example \textsc{RdnXor} has one bit of redundancy and one bit of synergy.  Yet for this example, $\WMS( \setX : Y ) = 0$ bits.}
	\label{fig:RS}
\end{figure}

Representing eq.~\eqref{eq:chechik1} for $n=2$ as a PI-diagram (\figref{fig:PID2_chechik}) reveals that $\WMS$ is the synergy between $X_1$ and $X_2$ \emph{minus} their redundancy.  Thus, when there is an equal magnitude of synergy and redundancy between $X_1$ and $X_2$ (as in \textsc{RdnXor}, \figref{fig:RS}), WholeMinusSum synergy is \emph{zero}---leading one to \emph{erroneously} conclude there is no synergy or redundancy present.\footnote{This is deeper than \cite{berry03}'s point that a mish-mash of synergy and redundancy across different states of $y \in Y$ can average to zero.  \figref{fig:RS} evaluates to zero for \emph{every state} $y \in Y$.}  

The PI-diagram for $n=3$ (\figref{fig:PID3_chechik}) reaveals that WholeMinusSum double-subtracts PI-regions \{1,2\}, \{1,3\}, \{2,3\} and triple-subtracts PI-region \{1,2,3\}, revealing that for $n>2$ $\WMS\left( \setX : Y\right)$ becomes synergy minus the redundancy \emph{counted multiple times}.

A concrete example demonstrating WholeMinusSum's ``synergy minus redundancy'' behavior is \textsc{RdnXor} (\figref{fig:RS}) which overlays examples \textsc{Rdn} and \textsc{Xor} to form a single system.  The target $Y$ has two bits of uncertainty, i.e. $\ent{Y}=2$.  Like \textsc{Rdn}, either $X_1$ or $X_2$ identically specifies the letter of $Y$ (\bin{r}/\bin{R}), making one bit of redundant information.  Like \textsc{Xor}, only the coalition $X_1 X_2$ specifies the digit of $Y$ (\bin{0}/\bin{1}), making one bit of synergistic information.  Together this makes one bit of redundancy and one bit of synergy.

Note that in \textsc{RdnXor} every state $y \in Y$ conveys one bit of redundant information and one bit of synergistic information, e.g. for the state $y=\bin{r0}$ the letter ``\bin{r}'' is specified redundantly and the digit ``\bin{0}'' is specified synergistically.  Example \textsc{RdnUnqXor} (Appendix \ref{app:extra}) extends \textsc{RdnXor} to demonstrate redundant, unique, and synergistic information for every state $y \in Y$.

In summary, WholeMinusSum \emph{underestimates} synergy for all $n$ with the potential gap increasing with $n$.  Equivalently, we say that WholeMinusSum synergy is a \emph{lowerbound} on the intuitive synergy with the bound becoming looser with $n$.  

\subsection{Correlational importance: $\Delta \opI \left( \setX ; Y \right)$}
\label{sect:deltaI}
Correlational importance, denoted $\Delta \opname{I}$, comes from \cite{panzeri99, nirenberg01, nirenberg03, pola03, latham-05}.  Correlational importance quantifies the ``informational importance of conditional dependence'' or the ``information lost when ignoring conditional dependence'' among the predictors decoding target $Y$.  As conditional dependence is necessary for synergy, $\Delta \opI$ seems related to our intuitive conception of synergy.  $\Delta \opI$ is defined as,

\begin{eqnarray}
\label{eq:delta0}
	\Delta \opname{I}\left( \setX ; Y\right) &\equiv& \DKL{\Prob{Y|\X}} { {\textstyle \Pr_{\ind}}\left(Y \middle|\setX\right)  } \\
\label{eq:delta1} &=& \sum_{y, \mathbf{x} \in Y, \setX} \Prob{y, x_{1\ldots n}} \log \frac{\Prob{y|x_{1 \ldots n}}}{\Pr_{\ind}(y|\mathbf{x}) } \; ,
\end{eqnarray}

where ${\textstyle \Pr_{\textnormal{ind}}}\left( y | \setx \right) \equiv   \frac{\Prob{y} \prod_{i=1}^n \Prob{x_i|y}}{ \sum_{y^\prime} \Prob{y^\prime} \prod_{i=1}^n \Prob{x_i|y^\prime} }$.  After some algebra\footnote{See Appendix \ref{appendix:deltai} for the steps between eqs.~\eqref{eq:delta1} and \eqref{eq:deltaIlast}.} eq.~\eqref{eq:delta1} becomes,

\begin{equation}
\label{eq:deltaIlast} 
	\Delta \opname{I}\left( \setX ; Y\right) = \opname{TC}\left( X_1; \cdots ; X_n \middle| Y \right) - \DKL{ \Prob{\X} }{ \sum_y \Prob{y} \prod_{i=1}^n \Prob{X_i | y} } \; .
\end{equation}

$\Delta \opI$ is conceptually innovative and moreover agrees with our intuition for all of our examples thus far.  Yet further examples reveal that $\Delta \opI$ measures something ever-so-subtly different from intuitive synergistic information.

The first example is \cite{berry03}'s Figure 4 where $\Delta \opI$ exceeds the mutual information $\info{\X}{Y}$ with $\Delta \opI \left(\setX ; Y\right)~=~ 0.0145$ and $\info{\X}{Y}~=~0.0140$.  This fact alone prevents interpreting $\Delta \opI$ as a loss of mutual information from $\info{\X}{Y}$.\footnote{Although $\Delta \opI$ can not be a loss of mutual information, it could still be a loss of some alternative information such as Wyner's common information \cite{lei-10}.}

\begin{figure}[h!bt]
	\centering
	\begin{minipage}[c]{0.3\linewidth} \centering \subfloat[$\Prob{x_1, x_2, y}$]{ \begin{tabular}{ c | c c } \cmidrule(r){1-2}
$\ \, X_1 \, X_2$  &$Y$ \\
\cmidrule(r){1-2} 
\bin{0 0} & \bin{0} & \quad \nicefrac{1}{4}\\
\bin{0 1} & \bin{0} & \quad \nicefrac{1}{4}\\
\bin{1 0} & \bin{0} & \quad \nicefrac{1}{4}\\
\bin{1 1} & \bin{1} & \quad \nicefrac{1}{4}\\
\cmidrule(r){1-2} 
\end{tabular} } \end{minipage}	
    \begin{minipage}[c]{0.33\linewidth} \centering
	\subfloat[PI-diagram]{ \includegraphics[width=1.15in]{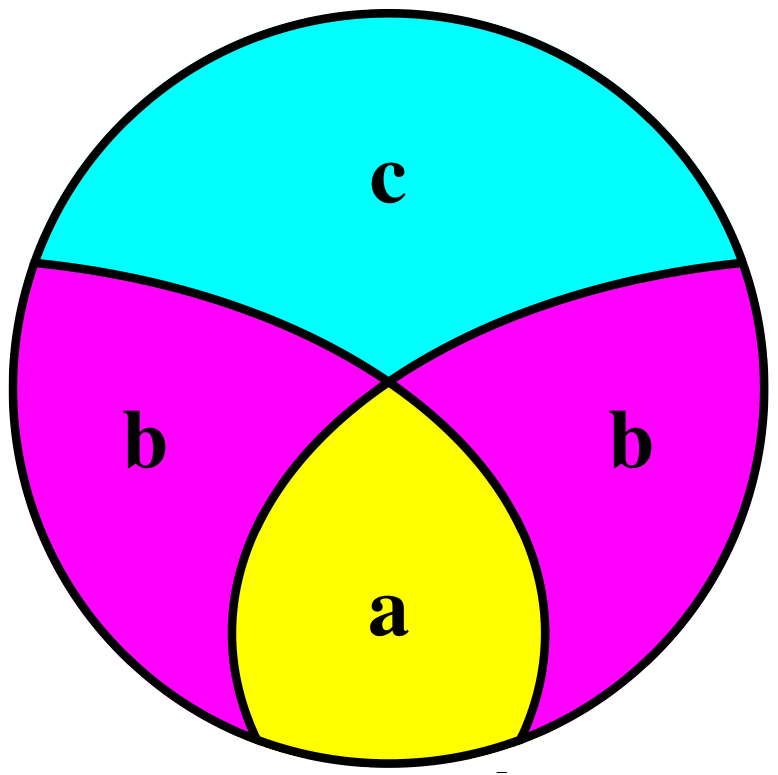} }
	\end{minipage}
	\begin{minipage}[c]{0.33\linewidth} \centering \begin{eqnarray*}
	0.189 \leq &c& \leq 0.5 \\
	0 \leq &b& \leq 0.311 \\
	0 \leq &a& \leq 0.311 \\
\end{eqnarray*}
\end{minipage}
	\subfloat[circuit diagram]{ \includegraphics[width=2.2in]{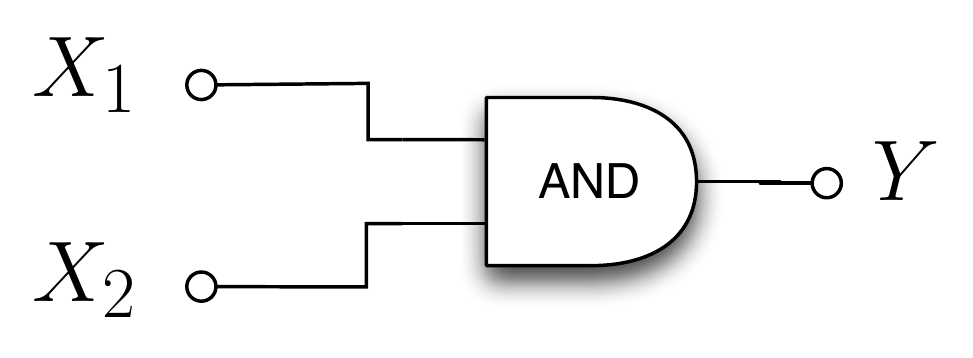} }
	\caption{Example \textsc{And}.  The exact PI-decomposition of an AND-gate remains uncertain.  But we can bound $a$, $b$, and $c$ using $\WMS$ and $\mathcal{S}_{\max}$.  In section \ref{sect:synergy} these bounds will be tightened.  Most intriguingly, we'll show that $a > 0$ despite $\info{X_1}{X_2}=0$. }
	\label{fig:AND}
\end{figure}

Could $\Delta \opI$ upperbound synergy instead?  We turn to example \textsc{And} (\figref{fig:AND}) with $n=2$ independent binary predictors and target $Y$ is the AND of $X_1$ and $X_2$.  Although \textsc{And}'s PI-region exact decomposition remains uncertain, we can still bound the synergy.  For example \textsc{And}, the $\WMS(\{X_1, X_2\} : Y ) \approx 0.189$ and $\S_{\max}\left( \{X_1, X_2\} : Y \right) = 0.5$ bits.  So we know the synergy must be between $(0.189, 0.5]$ bits.  Despite this, $\Delta \opI \left( \setX ; Y \right) = 0.104$ bits, thus $\Delta \opI$ does not upperbound synergy.  

Finally, in the face of duplicate predictors $\Delta \opI$ often \emph{decreases}.  From example \textsc{And} to \textsc{AndDuplicate} (Appendix \ref{sect:anddup}, \figref{fig:AndDup}) $\Delta \opI$ drops $63\%$ to 0.038 bits.

Taking all three examples together, we conclude $\Delta \opI$ measures something fundamentally different from synergistic information.

\section{Synergistic mutual information}
\label{sect:synergy}
We are all familiar with the English expression describing synergy as when the whole exceeds the ``sum of its parts''.  Although this informal adage captures the intuition underlying synergy, the formalization of this adage, WholeMinusSum synergy, ``double-counts'' whenever there is duplication (redundancy) among the parts.  A mathematically correct adage should change ``sum'' to ``union''---meaning synergy occurs when the whole exceeds the \emph{union} of its parts.  The sum adds duplicate information multiple times, whereas the union adds duplicate information only once.  The union of parts never exceeds the sum.

The guiding intuition of ``whole minus union'' leads us to a novel measure denoted $\Svkk{ \{ X_1, \ldots, X_n \}}{Y}$, or $\Svkk{\setX}{Y}$, as the mutual information in the whole beyond the union of elements $\{X_1, \ldots, X_n\}$.

Unfortunately, there's no established measure of ``union-information'' in contemporary information theory.  We introduce a novel technique, inspired by \cite{maurer99}, for defining the union information among $n$ predictors.  We numerically compute the union information by noisifying the joint distribution $\Prob{\X \middle| Y}$ such that only the correlations with singleton predictors are preserved.  This is achieved like so,

\begin{equation}
	\label{eq:ydaggerdef}
	\Ivkk{\{X_1, \ldots, X_n\}}{Y} \equiv {\displaystyle \min_{\Probstar{X_1, \ldots, X_n, Y}}} \opI^*\!\left( \X : Y \right)
\end{equation}
\[
\hspace{2.1in} \textnormal{subject to: } \ \Probstar{X_i, Y} = \Prob{X_i,Y} \ \forall i,
\]

where $\opI^*\!\left( \X : Y \right) \equiv \DKL{ \Probstar{\X, Y} }{ \Probstar{\X} \Probstar{Y} }$.

Without any constraint on the distribution $\Probstar{X_1, \ldots, X_n, Y}$, the minimum of eq.~\eqref{eq:ydaggerdef} is trivially found to be zero bits because simply setting $\Probstar{\X}$ to a constant makes $\opI^*(\X~:~Y)=0$ bits.  Therefore we must put some constraint on $\Probstar{X_1, \ldots, X_n, Y}$.  As all bits a singleton $X_i$ knows about $Y$ are determined by the joint distribution $\Prob{X_i,Y}$, we simply prevent the minimization from altering these distributions, and presto we arrive at the constraint $\Probstar{X_i,Y}~=~\Prob{X_i,Y}~\forall i$.\footnote{We could have instead chosen the \emph{looser} constraint $\opI^{*}(X_i:Y) = \info{X_i}{Y} \ \forall i$, but $\Probstar{X_i,Y} = \Prob{X_i,Y} \ \forall i$ ensures we preserve the ``same bits'', not just the same magnitude of bits.}  Finally, we prove that a minimum of eq.~\eqref{eq:ydaggerdef} always exists because setting  $\Probstar{x_1, \ldots, x_n, y} = \Prob{y} \prod_{i=1}^n \Prob{x_i|y}$ always satisfies the constraints.

\begin{figure}[h!bt]
	\centering
	\begin{minipage}[c]{0.3\linewidth} \centering \subfloat[$\Probstar{x_1, x_2, y}$]{ \begin{tabular}{ c | c c } \cmidrule(r){1-2}
$\ \, X_1 \, X_2$  &$Y$ \\
\cmidrule(r){1-2} 
\bin{0 0} & \bin{0} & \quad \nicefrac{1}{3}\\
\bin{0 1} & \bin{0} & \quad \nicefrac{1}{6}\\
\bin{1 0} & \bin{0} & \quad \nicefrac{1}{6}\\
\bin{1 1} & \bin{0} & \quad \nicefrac{1}{12}\\
\bin{1 1} & \bin{1} & \quad \nicefrac{1}{4}\\
\cmidrule(r){1-2} 
\end{tabular} } \end{minipage}	
    \begin{minipage}[c]{0.33\linewidth} \centering
	\subfloat[PI-diagram]{ \includegraphics[width=1.15in]{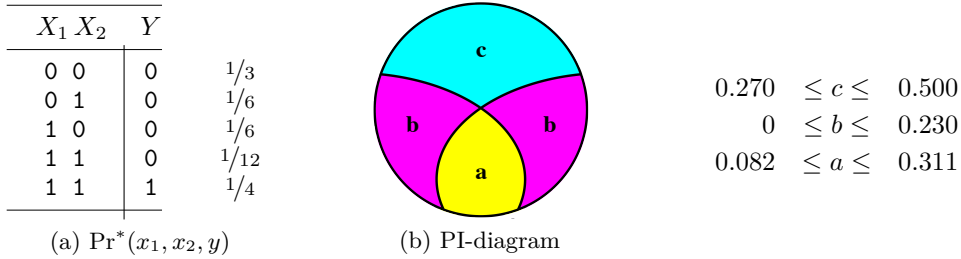} }
	\end{minipage}
	\begin{minipage}[c]{0.33\linewidth} \centering \begin{eqnarray*}
0.270 &\leq c \leq& 0.500 \\
0 &\leq b \leq& 0.230 \\ 
0.082 &\leq a \leq& 0.311 \\
\end{eqnarray*}
\end{minipage}	
	\caption{Revisiting example \textsc{And}.  Using the analytic upperbound on $\Ivk$ in Appendix \ref{appendix:ivkupperbound}, we arrive at the $\textstyle{\Pr^*}$ distribution in (a).  Using this distribution, we tighten the bounds on $a$, $b$, and $c$.  Intriguingly, we see that despite $\info{X_1}{X_2}=0$, that $a > 0$.  \textbf{Note}: Previous versions (preprints) of this paper \emph{erroneously asserted} independent predictors could not convey redundant information, i.e. that $\info{X_1}{X_2}=0$ entailed $\Icape{X_1,X_2}{Y}=0$.}
	\label{fig:ANDrevisited}
\end{figure}

Unfortunately, we currently have no analytic way to calculate eq.~\eqref{eq:ydaggerdef}, however, we do have an analytic upperbound on it.  Applying this to \textsc{And}'s PI-decomposition allows us to tighten the bounds in \figref{fig:AND} to those in \figref{fig:ANDrevisited}.

Our union-information measure $\Ivk$ satisfies several desired properties for a union-information measure.\footnote{For details see Section \ref{section:desiredprops} and Appendix \ref{appendix:axioms}.}  Once the union information is computed, the $\Svk$ synergy is simply,

\begin{equation}
\label{eq:synergydef}	\Svkk{ \{X_1, \ldots, X_n\}}{Y} \equiv \info{\X}{Y} - \Ivkk{\{X_1, \ldots, X_n\}}{Y} \; .
\end{equation}

$\Svk$ synergy quantifies the total ``informational work'' strictly the coalitions within $\X$ perform in reducing the uncertainty of $Y$.  Pleasingly, $\Svk$ is bounded\footnote{Proven in Appendix \ref{appendix:boundsproof}.} by the WholeMinusSum synergy (which underestimates the intuitive synergy) and $\S_{\max}$ (which overestimates intuitive synergy),
\begin{equation}
	\label{eq:allbounds}
	\max \left[ 0, \WMS\left( \setX : Y \right) \right] \leq \Svkk{\setX}{Y} \leq \S_{\max}\left( \setX : Y \right) \leq \info{\X}{Y} \; .
\end{equation}


\section{Properties of $\IcupVK$}
\label{section:desiredprops}
Our measure of the union information $\IcupVK$ satisfies several desirable properties for the union-information\footnote{For proofs see Appendix~\ref{appendix:axioms}.}:

\begin{enumerate}
    \item[\textbf{(GP)}] Global Positivity. $\IcupVKK{\setX}{Y} \geq 0$

    \item[\textbf{(SR)}] Self-Redundancy.  The union information a single predictor $X_1$ has about the target $Y$ is equal to the Shannon mutual information between the predictor and the target, i.e. $\IcupVKK{X_1}{Y} = \info{X_1}{Y}$.
    
    \item[$\mathbf{(S_0)}$] Weak Symmetry.  $\IcupVKK{X_1, \ldots, X_n}{Y}$ is invariant under reordering $X_1, \ldots, X_n$.

    \item[$\mathbf{(M)}$] Monotonicity.  $\IcupVKK{X_1, \ldots, X_{n}}{Y} \leq \IcupVKK{X_1, \ldots, X_n, W}{Y}$  with equality if $W$ is ``informationally poorer'' than some $X_i \in \{X_1, \ldots, X_{n}\}$, i.e. $\exists \ \ent{W|X_i}=0$ for some $i \in \{1, \ldots, n\}$.
    
    \item[\textbf{(TM)}] Target Monotonicity.  For all random variables $Y$ and $Z$, $\IcupVKK{ \setX }{Y} \leq \IcupVKK{\setX}{YZ}$.

    \item[$\mathbf{(LP_0)}$] Weak Local Positivity.  For $n=2$ predictors, the derived ``partial informations'' \cite{plw-10} are nonnegative.  This is equivalent to, 
\begin{equation*}
    \max\left[ \info{X_1}{Y}, \info{X_2}{Y} \right] \leq \IcupVKK{X_1,X_2}{Y} \leq \info{X_1 X_2}{Y} \; .
\end{equation*}

    \item[$\mathbf{(Id_1)}$] Strong Identity. $\IcupVKK{X_1, \ldots, X_n}{\X} = \ent{\X}$.
\end{enumerate}

\section{Applying the measures to our examples}
\label{sect:results}

Table \ref{fig:thetable} summarizes the results of all four measures applied to our examples.

\textsc{Rdn} (\figref{fig:exampleR}).  There is exactly one bit of redundant information and all measures reach their intended answer.  For the axiomatically minded, the equality condition of $\mathbf{(M)}$ is sufficient for the desired answer.

\textsc{Unq} (\figref{fig:exampleU}).  $\S_{\max}$'s miscategorization of unique information as synergistic reveals itself.    Intuitively, there are two bits of unique information and no synergy.  However, $\S_{\max}$ reports one bit of synergistic information.  For the axiomatically minded, property (\textbf{Id}) is sufficient (but not nessecary) for the desired answer.

\textsc{Xor} (\figref{fig:xor}).  There is exactly one bit of synergistic information.  All measures reach the desired answer of 1 bit.

\textsc{XorDuplicate} (\figref{fig:XorDup}).  Target $Y$ is specified by the coalition $X_1 X_2$ as well as by the coalition $X_3 X_2$, thus $\info{X_1 X_2}{Y} = \info{X_3 X_2}{Y} = \ent{Y} = 1$ bit.  All measures reach the expected answer of 1 bit.

\textsc{XorLoses} (\figref{fig:xorallunique}).  Target $Y$ is specified by the coalition $X_1 X_2$ as well as by the singleton $X_3$, thus $\info{X_1 X_2}{Y} = \info{X_3}{Y} = \ent{Y} = 1$ bit.  Together this means there is one bit of redundancy between the coalition $X_1 X_2$ and the singleton $X_3$ as illustrated by the $+1$ in PI-region $\{3,12\}$.  All measures account for this redundancy and reach the desired answer of 0 bits.

\textsc{RdnXor} (\figref{fig:RS}).  This example has one bit of synergy as well as one bit of redundancy.  In accordance with \figref{fig:PID2_chechik}, WholeMinusSum measures \emph{synergy minus redundancy} to calculate $1-1=0$ bits.  On the other hand, $\S_{\max}$, $\Delta \opname{I}$, and $\Svk$ are not mislead by the co-existance of synergy and redundancy and correctly report 1 bit of synergistic information.

\textsc{And} (\figref{fig:AND}).  This example is a simple case where correlational importance, $\Delta \opI ( \setX ; Y )$, disagrees with the intuitive value for synergy.  The WholeMinusSum synergy---an unambiguous \emph{lowerbound} on the intuitive synergy---is $0.189$ bits, yet $\Delta \opI \left( \setX ; Y \right) = 0.104$ bits.  We can't perfectly determine $\Svk$, but we can lowerbound $\Svk$ using our analytic bound, as well as upperbound it using $\S_{\max}$.  This gives $0.270 \leq \Svk \leq \nicefrac{1}{2}$.

The three supplementary examples in Appendix \ref{app:extra}: \textsc{RdnUnqXor}, \textsc{AndDuplicate}, and \textsc{XorMultiCoal} aren't essential for understanding this paper and are for the intellectual pleasure of advanced readers.

Table \ref{fig:thetable} shows that no prior measure of synergy consistently matches intuition even for $n=2$.   To summarize,
\begin{enumerate}
	\item	$\opI_{\max}$ synergy, $\S_{\max}$, overestimates the intuitive synergy when two or more predictors convey unique information about the target (e.g. \textsc{Unq}).
	\item	WholeMinusSum synergy, $\WMS$, inadvertently double-subtracts redundancies and thus underestimates the intuitive synergy (e.g. \textsc{RdnXor}).  Duplicating predictors often decreases WholeMinusSum synergy (e.g. \textsc{AndDuplicate}).
	\item	Correlational importance, $\Delta \opI$, is not bounded by the Shannon mutual information, underestimates the known lowerbound on synergy (e.g. \textsc{And}), and duplicating predictors often decreases correlational importance (e.g. \textsc{AndDuplicate}).  Altogether, $\Delta \opI$ does not quantify the intuitive  synergistic information (nor was it intended to).
\end{enumerate}


\begin{table}[tb]
\centering
	\begin{tabular}{  l l l l l l } \toprule
 \addlinespace
		Example & $\S_{\max}$ & $\WMS$ & $\Delta \opname{I}$ & $\Svk$\\
	\midrule
\textsc{Rdn} & 0 & --1 & 0 & 0 \\
\textsc{Unq} & \red{1} & 0 & 0 & 0 \\
\textsc{Xor} & 1 & 1 & 1 & 1 \\
\addlinespace
\textsc{XorDuplicate} & 1 & 1 & 1 & 1 \\
\textsc{XorLoses} & 0 & 0 & 0 & 0 \\
\addlinespace
\textsc{RdnXor} & 1 & \red{0} & 1 & 1 \\
\textsc{And} & $\nicefrac{1}{2}$ & \red{0.189} & \red{0.104} & [0.270,$\nicefrac{1}{2}$] \\

\addlinespace
\textsc{RdnUnqXor} & \red{2} & \red{0} & 1 & 1 \\
\textsc{AndDuplicate} & $\nicefrac{1}{2}$ & \red{--0.123} & \red{0.038} & [0.270,$\nicefrac{1}{2}$] \\
\textsc{XorMultiCoal} & 1 & 1 & 1 & 1 \\
\bottomrule
\end{tabular}
\caption{Synergy measures for our examples.  Answers conflicting with intuitive synergistic information are in \!\!{\red{red}}.  The $\Svk$ value for \textsc{And} and \textsc{AndDuplicate} is not conclusively known, but can be bounded.}
\label{fig:thetable}
\end{table}

\section{Conclusion}
Fundamentally, we assert that synergy quantifies how much the whole exceeds the \emph{union} of its parts.  Considering synergy as the whole minus the \emph{sum} of its parts inadvertently ``double-subtracts'' redundancies, thus \emph{underestimating} synergy.  Within information theory, PI-diagrams, a generalization of Venn diagrams, are immensely helpful in improving one's intuition for synergy.

We demonstrated with \textsc{RdnXor} and \textsc{RdnUnqXor} that a single state can simultaneously carry redundant, unique, and synergistic information.  This fact is underappreciated, and prior work often implicitly assumed these three types of information could not coexist in a single state.

We introduced a novel measure of synergy, $\Svk$, (eq.~\eqref{eq:synergydef}).  Unfortunately our expression is not easily computable, and until we have an explicit analytic solution to the minimization in $\Ivk$ the best one can do is numerical optimization using our analytic upperbound (Appendix \ref{appendix:ivkupperbound}) as a starting point.

Along with our examples, we consider our introduction of a candidate for the union information, $\Ivk$ (eq.~\eqref{eq:ydaggerdef}) and its upperbound our primary contributions to the literature.

Finally, by means of our analytic upperbound on $\Ivk$ we've shown that, at least for our measure, \emph{independent predictors can convey redundant information about a target}, e.g. \figref{fig:ANDrevisited}.

\subsubsection*{Acknowledgments}
We thank Suzannah Fraker, Tracey Ho, Artemy Kolchinsky, Chris Adami, Giulio Tononi, Jim Beck, Nihat Ay, and Paul Williams for extensive discussions.  This research was funded by the Paul G. Allen Family Foundation and a DOE CSGF fellowship to VG.

\bibliographystyle{apalike}
\bibliography{ref1}

\clearpage
\appendix

\section{Three extra examples}
\label{app:extra}
For the reader's intellectual pleasure, we include three more sophisticated examples: \textsc{RdnUnqXor}, \textsc{AndDuplicate}, and \textsc{XorMultiCoal}.

\begin{figure}[h!tb]
	\centering
\begin{minipage}[b]{0.45\linewidth} \centering \begin{tabular}{ c | c c } \cmidrule(r){1-2}
$X_1 \ \  X_2$  &$Y$ \\
\cmidrule(r){1-2} 
\bin{ra0 rb0} & \bin{rab0} & \quad \nicefrac{1}{32}\\
\bin{ra0 rb1} & \bin{rab1} & \quad \nicefrac{1}{32}\\
\bin{ra1 rb0} & \bin{rab1} & \quad \nicefrac{1}{32}\\
\bin{ra1 rb1} & \bin{rab0} & \quad \nicefrac{1}{32}\\
\addlinespace
\bin{ra0 rB0} & \bin{raB0} & \quad \nicefrac{1}{32}\\
\bin{ra0 rB1} & \bin{raB1} & \quad \nicefrac{1}{32}\\
\bin{ra1 rB0} & \bin{raB1} & \quad \nicefrac{1}{32}\\
\bin{ra1 rB1} & \bin{raB0} & \quad \nicefrac{1}{32}\\
\addlinespace
\bin{rA0 rb0} & \bin{rAb0} & \quad \nicefrac{1}{32}\\
\bin{rA0 rb1} & \bin{rAb1} & \quad \nicefrac{1}{32}\\
\bin{rA1 rb0} & \bin{rAb1} & \quad \nicefrac{1}{32}\\
\bin{rA1 rb1} & \bin{rAb0} & \quad \nicefrac{1}{32}\\
\addlinespace
\bin{rA0 rB0} & \bin{rAB0} & \quad \nicefrac{1}{32}\\
\bin{rA0 rB1} & \bin{rAB1} & \quad \nicefrac{1}{32}\\
\bin{rA1 rB0} & \bin{rAB1} & \quad \nicefrac{1}{32}\\
\bin{rA1 rB1} & \bin{rAB0} & \quad \nicefrac{1}{32}\\
\cmidrule(r){1-2} 
\end{tabular} \end{minipage}
\begin{minipage}[b]{0.45\linewidth} \centering \begin{tabular}{ c | c c } \cmidrule(r){1-2}
$X_1 \ \  X_2$  &$Y$ \\
\cmidrule(r){1-2} 
\bin{Ra0 Rb0} & \bin{Rab0} & \quad \nicefrac{1}{32}\\
\bin{Ra0 Rb1} & \bin{Rab1} & \quad \nicefrac{1}{32}\\
\bin{Ra1 Rb0} & \bin{Rab1} & \quad \nicefrac{1}{32}\\
\bin{Ra1 Rb1} & \bin{Rab0} & \quad \nicefrac{1}{32}\\
\addlinespace
\bin{Ra0 RB0} & \bin{RaB0} & \quad \nicefrac{1}{32}\\
\bin{Ra0 RB1} & \bin{RaB1} & \quad \nicefrac{1}{32}\\
\bin{Ra1 RB0} & \bin{RaB1} & \quad \nicefrac{1}{32}\\
\bin{Ra1 RB1} & \bin{RaB0} & \quad \nicefrac{1}{32}\\
\addlinespace
\bin{RA0 Rb0} & \bin{RAb0} & \quad \nicefrac{1}{32}\\
\bin{RA0 Rb1} & \bin{RAb1} & \quad \nicefrac{1}{32}\\
\bin{RA1 Rb0} & \bin{RAb1} & \quad \nicefrac{1}{32}\\
\bin{RA1 Rb1} & \bin{RAb0} & \quad \nicefrac{1}{32}\\
\addlinespace
\bin{RA0 RB0} & \bin{RAB0} & \quad \nicefrac{1}{32}\\
\bin{RA0 RB1} & \bin{RAB1} & \quad \nicefrac{1}{32}\\
\bin{RA1 RB0} & \bin{RAB1} & \quad \nicefrac{1}{32}\\
\bin{RA1 RB1} & \bin{RAB0} & \quad \nicefrac{1}{32}\\
\cmidrule(r){1-2} 
\end{tabular} \end{minipage}
\caption*{(a) $\Prob{x_1, x_2, y}$}
	\setcounter{subfigure}{1}
	\subfloat[circuit diagram]{ \includegraphics[height=1.6in]{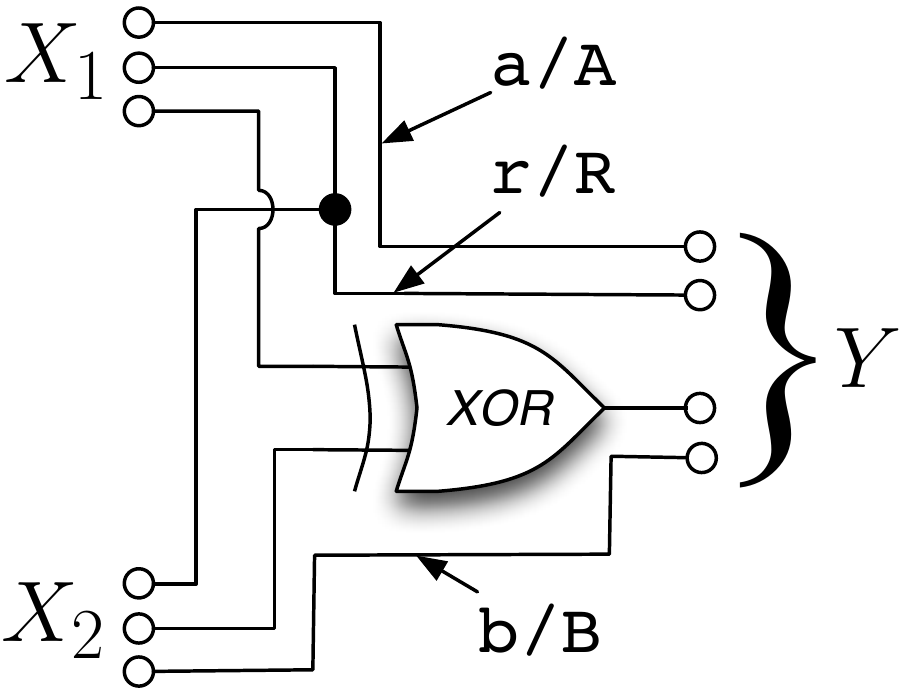} }	
	\subfloat[PI-diagram]{ \includegraphics[height=1.4in]{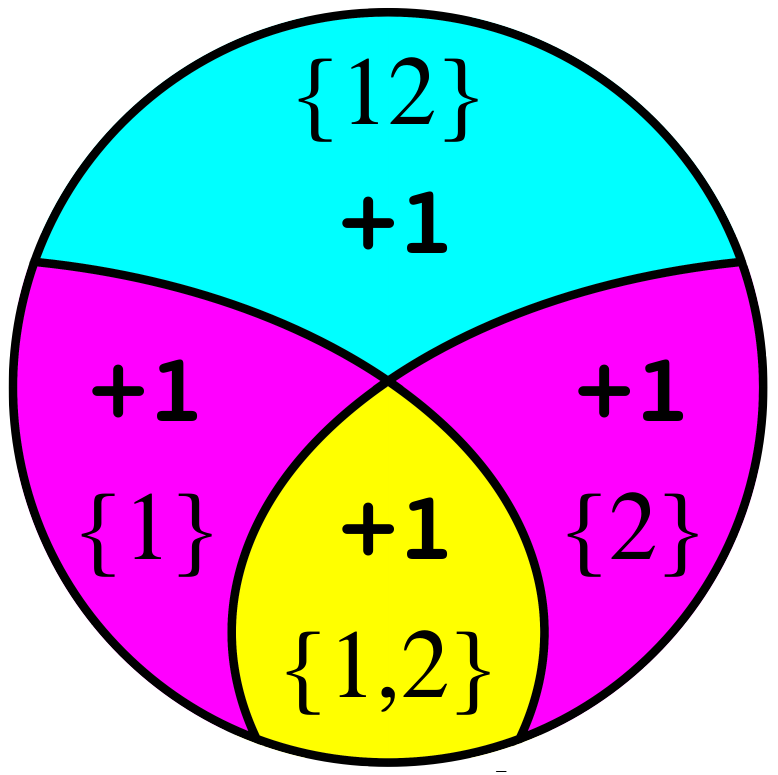} }
	\caption{Example \textsc{RdnUnqXor} weaves examples \textsc{Rdn}, \textsc{Unq}, and \textsc{Xor} into one.  $\info{X_1 X_2}{Y}~=~\ent{Y}~=~4$ bits.  This example is pleasing because it puts exactly one bit in each PI-region.}
	\label{fig:quad}
\end{figure}

\subsubsection{Example AndDuplicate}
\label{sect:anddup}
\textsc{AndDuplicate} adds a duplicate predictor to example \textsc{And} to show how $\Delta \opI$ responds to a duplicate predictor in a less pristine example than \textsc{Xor}. Unlike \textsc{Xor}, in example \textsc{And} there's also unique and redundant information.  Will this cause the loss of synergy in the spirit of \textsc{XorLoses}?  Taking each one at a time:

\begin{itemize}
	\item Predictor $X_2$ is unaltered from example \textsc{And}.  Thus $X_2$'s unique information stays the same.  \textsc{And}'s $\{2\} \to $ \textsc{AndDuplicate}'s $\{2\}$.
	\item Predictor $X_3$ is identical to $X_1$.  Thus all of $X_1$'s unique information in \textsc{And} becomes redundant information between predictors $X_1$ and $X_3$.  \textsc{And}'s $\{1\} \to $ \textsc{AndDuplicate}'s $\{1,3\}$.
	\item In \textsc{And} there is synergy between $X_1$ and $X_2$, and this synergy is still present in \textsc{AndDuplicate}.  Just as in \textsc{XorDuplicate}, the only difference is that now an identical synergy also exists between $X_3$ and $X_2$.  Thus \textsc{And}'s $\{12\} \to $ \textsc{AndDuplicate}'s $\{12,23\}$.
	\item Predictor $X_3$ is identical to $X_1$.  Therefore any information in \textsc{And} that is specified by both $X_1$ and $X_2$ is now specified by $X_1$, $X_2$, and $X_3$.  Thus \textsc{And}'s $\{1,2\} \to $ \textsc{AndDuplicate's} $\{1,2,3\}$.
\end{itemize}

\begin{figure}[h!bt]
	\centering
	\begin{minipage}[c]{0.4\linewidth} \centering	
	\subfloat[$\Prob{x_1, x_2, x_3, y}$]{ \begin{tabular}{ c | c c } \cmidrule(r){1-2}
$\ \, X_1 \, X_2 \, X_3$  &$Y$ \\
\cmidrule(r){1-2} 
\bin{0 0 0} & \bin{0} & \quad \nicefrac{1}{4}\\
\bin{0 1 0} & \bin{0} & \quad \nicefrac{1}{4}\\
\bin{1 0 1} & \bin{0} & \quad \nicefrac{1}{4}\\
\bin{1 1 1} & \bin{1} & \quad \nicefrac{1}{4}\\
\cmidrule(r){1-2} 
\end{tabular} }
\end{minipage} 	\begin{minipage}[c]{0.55\linewidth} \centering
	\subfloat[circuit diagram]{ \includegraphics[width=1.9in]{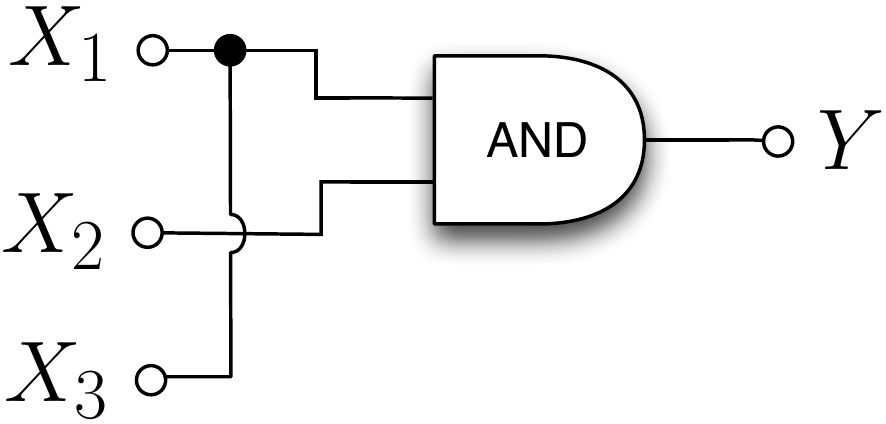} }
	\end{minipage}
\begin{minipage}[c]{\linewidth} \centering
	\subfloat[PI-diagram]{ \includegraphics[width=4.6in]{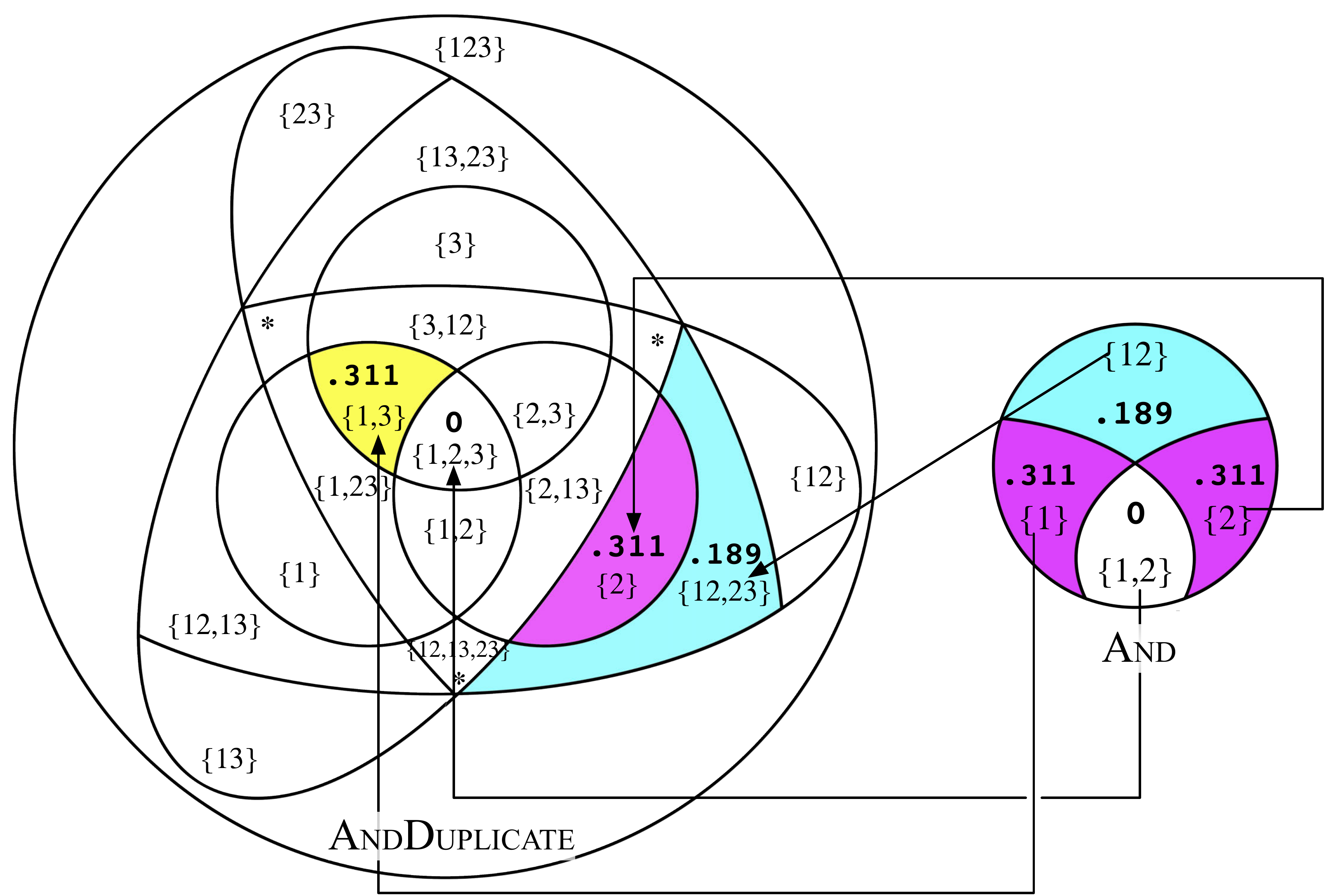} \label{fig:AndDupC} }
	\end{minipage}
	\caption{Example \textsc{AndDuplicate}.  The total mutual information is the same as in \textsc{And}, $\info{X_1 X_2}{Y}=\info{X_1 X_2 X_3}{Y}=0.811$ bits.  Every PI-region in example \textsc{And} maps to a PI-region in \textsc{AndDuplicate}.  The intuitive synergistic information is unchanged from \textsc{And}.  However, correlational importance, $\Delta \opI$, arrives at $0.104$ bits of synergy for \textsc{And}, and $0.038$ bits for \textsc{AndDuplicate}.  $\Delta \opI$ is not invariant to duplicate predictors.}
	\label{fig:AndDup}
\end{figure}

\begin{figure}[h!bt]
	\centering
	\begin{minipage}[c]{0.4\linewidth} \centering
\subfloat[$\Prob{x_1,x_2,x_3,y}$]{ \begin{tabular}{ c | c c } \cmidrule(r){1-2}
$\ \;X_1 \ X_2 \ X_3$  &$Y$ \\
\cmidrule(r){1-2} 
\bin{ab ac bc} & \bin{0} & \quad \nicefrac{1}{8}\\
\bin{AB Ac Bc} & \bin{0} & \quad \nicefrac{1}{8}\\
\bin{Ab AC bC} & \bin{0} & \quad \nicefrac{1}{8}\\
\bin{aB aC BC} & \bin{0} & \quad \nicefrac{1}{8}\\
\addlinespace
\bin{Ab Ac bc} & \bin{1} & \quad \nicefrac{1}{8}\\
\bin{aB ac Bc} & \bin{1} & \quad \nicefrac{1}{8}\\
\bin{ab aC bC} & \bin{1} & \quad \nicefrac{1}{8}\\
\bin{AB AC BC} & \bin{1} & \quad \nicefrac{1}{8}\\
\cmidrule(r){1-2}
\end{tabular} }
\end{minipage} \begin{minipage}[c]{0.55\linewidth} \centering
	\subfloat[circuit diagram]{ \includegraphics[width=3.0in]{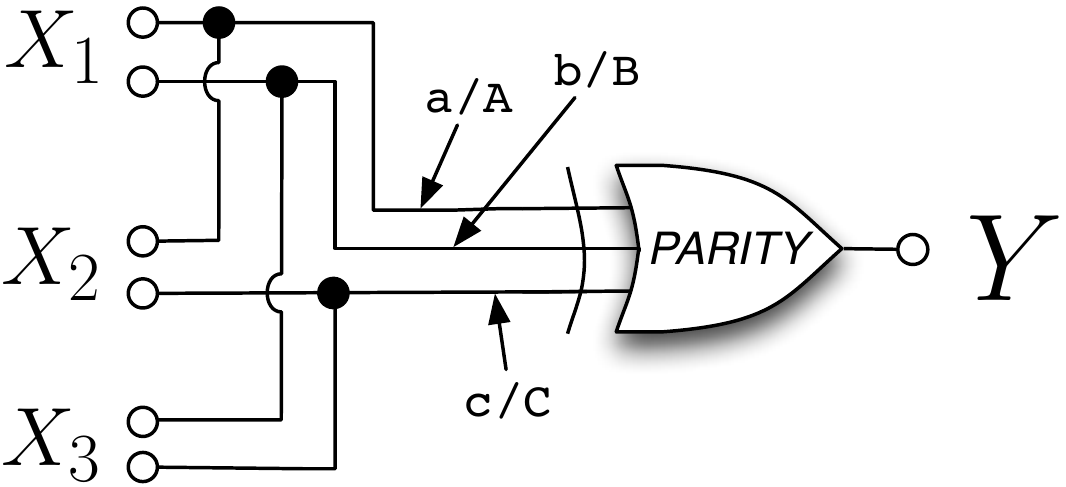} }
	\end{minipage}
\begin{minipage}[c]{0.55\linewidth} \centering
	\subfloat[PI-diagram]{ \includegraphics[height=2.5in]{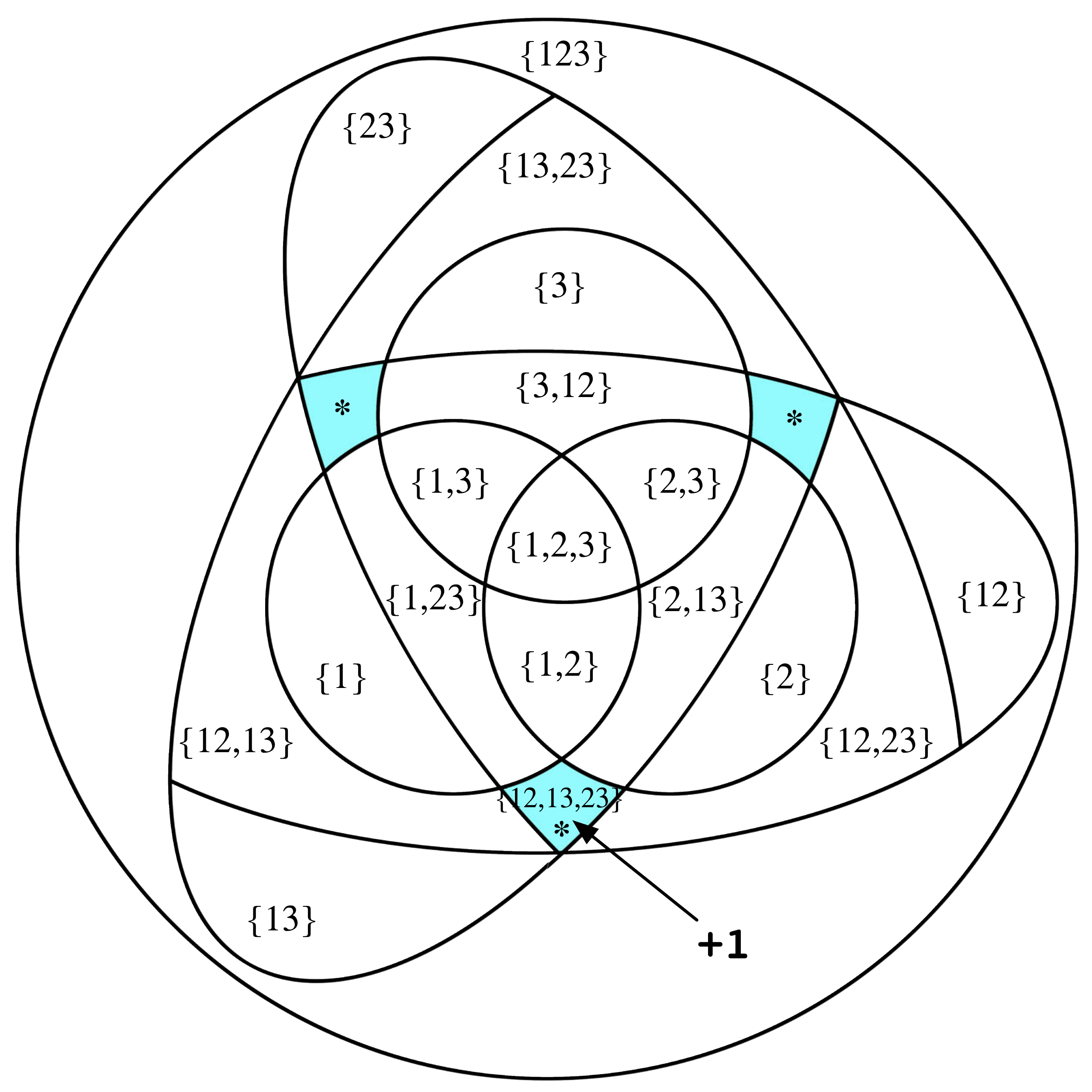} }
	\end{minipage}
	
	\caption{Example \textsc{XorMultiCoal} demonstrates how the same information can be specified by multiple coalitions.  In \textsc{XorMultiCoal} the target $Y$ has one bit of uncertainty, $\ent{Y} = 1$ bit, and $Y$ is the \emph{parity} of three incoming wires.  Just as the output of \textsc{Xor} is specified only after knowing the state of both inputs, the output of \textsc{XorMultiCoal} is specified only after knowing the state of all three wires.  Each predictor is distinct and has access to two of the three incoming wires.  For example, predictor $X_1$ has access to the \texttt{a/A} and \texttt{b/B} wires, $X_2$ has access to the \texttt{a/A} and \texttt{c/C} wires, and $X_3$ has access to the \texttt{b/B} and \texttt{c/C} wires.  Although no single predictor specifies $Y$, any coalition of two predictors has access to all three wires and fully specifies $Y$, $\info{X_1 X_2}{Y}~=~\info{X_1 X_3}{Y}~=~\info{X_2 X_3}{Y}~=~\ent{Y}~=~1$ bit.  In the PI-diagram this puts one bit in PI-region $\{12,13,23\}$ and zero everywhere else.  All measures reach the expected answer of 1 bit of synergy.}
	\label{fig:XorMulti}
\end{figure}

\clearpage
\section{Connecting back to $\Icap$}
Our candidate measure of the union information, $\IcupVK$, gives rise to a measure of the intersection-information denoted $\IcapVK$.  This is done by,

\begin{equation}
    \IcapVKK{ \setX }{Y} = \sum_{\mathbf{S} \subseteq \setX} (-1)^{|\mathbf{S}|+1} \IcupVKK{\mathbf{S}}{Y} \; .
\end{equation}

\section{Desired properties of $\Icup$}
What properties does $\IcapVK$ satisfy?  We originally worked on proofs for which properties $\IcapVK$ satisfies, but for $n>2$ we were blocked by not having an analytic solution to $\IcupVK$.  So we instead translated the $\Icap$ properties into the analogous $\Icup$ properties.  Although one can't always prove the $\Icap$ version from the analogous $\Icup$ property, it is a start.

\label{appendix:axioms}


    

    


In addition to the properties in Section \ref{section:desiredprops}, we 
We've proven that $\IcupVK$ \emph{does not satisfy} the property,
\begin{enumerate}
    \item[$\mathbf{(S_1)}$] Strong Symmetry. $\Icupe{X_1, \ldots, X_n}{Y}$ is invariant under reordering $X_1, \ldots, X_n, Y$.
\end{enumerate}

\subsubsection{Proof of \textbf{(GP)}}
Proven by the nonnegativity of mutual information.

\subsubsection{Proof of \textbf{(SR)}}
\begin{eqnarray*}
    \IcupVKK{X_1}{Y} &\equiv& \min_{ \substack{p^*(x_1,y) \\ p^*(x_1,y) = p(x_1,y)}} \infostar{X_1}{Y} \\
    &=& \info{X_1}{Y} \; .
\end{eqnarray*}


\subsubsection{Proof of $\mathbf{(S_0)}$}
There's only one instance of the terms in $\setX$ in the definition of $\IcupVK$, which is,
\begin{equation*}
    \IcupVKK{\setX}{Y} \equiv \VKbox \infostar{X_1 \cdots X_n}{Y} \; .
\end{equation*}

The term $\infostar{X_1 \cdots X_n}{Y}$ is invariant to the ordering of $X_1 \cdots X_n$.  This is due to $\Probstar{x_1, \ldots, x_n} = \Probstar{x_n, \ldots, x_1}$.  Thus $\IcupVK$ is invariant to the ordering of $\{X_1, \ldots, X_n\}$.

\subsubsection{Proof of $\mathbf{(LP_0)}$}
\begin{equation*}
    \IcupVKK{ \setX }{Y} \leq \info{\X}{Y} \; .
\end{equation*}

This is proven by the condition that $\Prob{X_1, \ldots, X_n,Y}$ satisfies the constraints on the minimizing distribution in $\IcupVK$.  Thus $\infostar{ \X }{Y} \leq \info{ \X }{Y}$.

\subsubsection{Disproof of $\mathbf{(S_1)}$}

We show that, $\IcupVKK{ \{X, Y\} }{Z} \not= \IcupVKK{ \{X,Z\} }{Y}$ by setting $X=Y$ where $\ent{X}~>~ 0$, and $Z$ is a constant, $\IcupVKK{ \{X, Y\} }{Z}=0$ yet $\IcupVKK{ \{X,Z\} }{Y}=\ent{X}$.

\subsubsection{Proof of $\mathbf{(Id_1)}$}

\begin{eqnarray}
	\Ivkk{ \setX }{ \X } &\equiv& \min_{\substack{p^*(X_1, \ldots, X_n, \X) \\ p^*(X_i,\X) = p(X_i, \X) \ \forall i}} \opI^*\left( \X : \X \right) \\
	&=& \min_{\substack{p^*(X_1, \ldots, X_n, \X) \\ p^*(X_i,\X) = p(X_i, \X) \ \forall i}} \opname{H^*}\!\left( \X \right) \; ,
\end{eqnarray}

    Then because $p^*(X_{1 \ldots n}) = p(X_{1 \ldots n})$,

\begin{equation}
    \Ivkk{ \setX }{ \X } = \ent{\X} \; .
\end{equation}





\section{Analytic upperbound on $\Ivkk{ \setX }{Y}$}
\label{appendix:ivkupperbound}

Our analytic upperbound on $\Ivk$ starts with the $n$ joint distributions we wish to preserve: $\Prob{X_1, Y}, \ldots , \Prob{X_n,Y}$.  From one these joint distributions, e.g. $\Prob{X_1,Y}$, we compute the marginal probability distribution $\Prob{Y}$ by summing over the index of $x_1 \in X_1$,

\begin{equation}
    \Prob{Y} = \left\{ \sum_{x_1 \in X_1} \Prob{x_1, y} : \forall y \in Y \right\} \; .
\end{equation}

Then, for every state $y \in Y$ we compute $n$ conditional distributions $\Prob{X_1|y}, \ldots, \Prob{X_n|y}$ via,

\begin{equation}
    \Prob{X_i|Y=y} = \left\{ \frac{\Prob{x_i,y}}{\Prob{y}} : \forall x_i \in X_i \right\} \; .
\end{equation}

With the marginal distribution $\Prob{Y}$ and the $|Y|\cdot n$ conditonal distributions, we construct a novel, artificial joint distribution $\Probstar{ X_1, \ldots, X_n, Y}$ defined by,

\begin{equation}
    \label{eq:probstar}
    \Probstar{x_1, \ldots, x_n, y} \equiv \Prob{y} \prod_{i=1}^n \Prob{x_i|y} \; .
\end{equation}

This novel, artificial joint distribution $\Probstar{X_1, \ldots, X_n, Y}$ satisfies the constraints $\Probstar{X_i,Y}=\Prob{X_i,Y}\ \forall i$.  This is proven by,

\begin{eqnarray}
    \Probstar{x_i,y} &=& \underbrace{\sum_{x_1 \in X_1} \cdots \sum_{x_n \in X_n}}_{\textnormal{All except $x_i \in X_i$}} \Probstar{x_1, \ldots, x_n, y} \\
    &=& \underbrace{\sum_{x_1 \in X_1} \cdots \sum_{x_n \in X_n}}_{\textnormal{All except $x_i \in X_i$}} \Prob{y} \prod_{j=1}^n \Prob{x_i|y} \\
    &=& \underbrace{\sum_{x_1 \in X_1} \cdots \sum_{x_n \in X_n}}_{\textnormal{All except $x_i \in X_i$}} \Prob{x_i,y} \prod_{\substack{j=1 \\ j \not= i}}^n \Prob{x_j|y} \\
    &=& \Prob{x_i,y} \underbrace{\underbrace{\sum_{x_1 \in X_1} \cdots \sum_{x_n \in X_n}}_{\textnormal{All except $x_i \in X_i$}} \prod_{\substack{j=1 \\ j \not= i}}^n \Prob{x_j|y}}_{\textnormal{sums to $1$}} \\
    &=& \Prob{x_i,y} \; .
\end{eqnarray}

\begin{figure}[h!t]
    \centering 
	\includegraphics[width=2in]{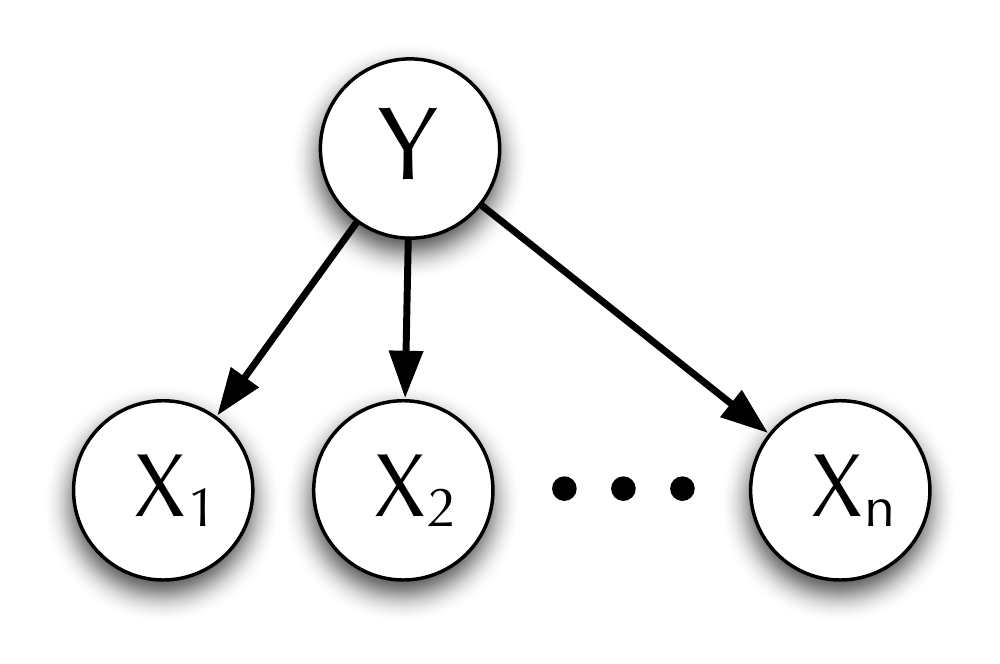} \caption{The Directed Acyclic Graph generating the joint distribution $\Probstar{x_1, \ldots, x_n, y}$.  This is a graphical representation of eq.~\eqref{eq:probstar}.}
	\label{fig:ProbstarDAG} 
\end{figure}

The upperbound on $\IcupVK$ is then the mutual information using this artificial $\textstyle{\Pr^*}$ distribution,

\begin{equation}
\opI^{*}\!\left( X_1 \ldots X_n : Y \right) =  \sum_{x_1 \in X_1} \!\cdots\! \sum_{x_n \in X_n} \sum_{y \in Y} \Probstar{x_1, \ldots, x_n, y} \log \frac{ \Probstar{x_1, \ldots, x_n, y} }{ \Probstar{x_1, \ldots, x_n} \Probstar{y}  } \; ,
\end{equation}

where the terms $\Probstar{x_1, \ldots, x_n}$ and $\Probstar{y}$ are defined by summing over the relevant indices of joint distribution $\Probstar{X_1, \ldots, X_n, Y}$,

\begin{eqnarray}
    \Probstar{x_1, \ldots, x_n} &=& \sum_{y^\prime \in Y} \Probstar{x_1, \ldots, x_n, y^\prime} \\
    &=& \sum_{y^\prime \in Y} \Prob{y^\prime} \prod_{i=1}^n \Prob{x_i|y^\prime} \; ;
\end{eqnarray}

\begin{eqnarray}
    \Probstar{y} &=& \sum_{x_1 \in X_1} \cdots \sum_{x_n \in X_n} \Probstar{x_1, \ldots, x_n, y} \\
    &=& \sum_{x_1 \in X_1} \cdots \sum_{x_n \in X_n} \Prob{y} \prod_{i=1}^n \Prob{x_i|y} \\
    &=& \Prob{y} \underbrace{\sum_{x_1 \in X_1} \cdots \sum_{x_n \in X_n} \prod_{i=1}^n \Prob{x_i|y}}_{\textnormal{sums to $1$}} \\
    &=& \Prob{y} \; .
\end{eqnarray}

Putting everything together, our analytic upperbound on $\IcupVK$ is,
\begin{eqnarray}
    \Ivkk{ \{X_1, \ldots, X_n\}}{Y} &\leq& \opI^*\!\left( \X : Y \right) \\
    &=& \sum_{x_1} \cdots \sum_{x_n} \sum_{y} \Probstar{x_1, \ldots, x_n, y} \log \frac{ \Probstar{x_1, \ldots, x_n, y} }{ \Probstar{x_1, \ldots, x_n} \Probstar{y}  } \\
    &=& \sum_{x_1} \cdots \sum_{x_n} \sum_{y} \Probstar{x_1, \ldots, x_n, y} \log \frac{\Prob{y} \prod_{i=1}^n \Prob{x_i|y}}{\Probstar{x_1, \ldots, x_n} \Prob{y}} \\
    &=& \sum_{x_1} \cdots \sum_{x_n} \sum_{y} \Probstar{x_1, \ldots, x_n, y} \log \frac{\prod_{i=1}^n \Prob{x_i|y}}{\Probstar{x_1, \ldots, x_n} } \\
\nonumber    &=& \sum_{y} \Prob{y} \sum_{x_1} \cdots \sum_{x_n} \prod_{i=1}^n \Prob{x_i|y} \log \frac{\prod_{i=1}^n \Prob{x_i|y}}{\sum_{y^\prime \in Y} \Prob{y^\prime} \prod_{i=1}^n \Prob{x_i|y^\prime} } \; .
\end{eqnarray}

\clearpage
\section{Essential proofs}
\label{appendix:proofs}
These proofs underpin essential claims about our introduced measure, synergistic mutual information.

\subsection{Proof duplicate predictors don't increase synergy}
We show that synergy being invariant to duplicate predictors follows from the equality condition of \textbf{(M)} of the intersection (as well as union) information.

We show that,
\begin{equation*}
    \Svkk{ \setX }{Y} = \Svkk{ \setX^\prime }{Y} \; ,
\end{equation*}
where $\setX^\prime \equiv \{X_1, \ldots, X_n, X_1\}$.  We show that $\Svkk{ \setX }{Y} - \Svkk{ \setX^\prime}{Y}=0$.

\begin{eqnarray}
0 &=& \Svkk{ \setX }{Y} - \Svkk{ \setX^\prime}{Y} \\
&=& \info{\X}{Y} - \IcupVKK{ \setX }{Y} - \info{\X X_1}{Y} + \IcupVKK{ \setX^\prime}{Y} \\
&=& \IcupVKK{ \setX^\prime}{Y} - \IcupVKK{ \setX }{Y}\\
&=& \sum_{\setT \subseteq \setX^\prime} (-1)^{|\mathbf{T}|+1} \IcapVKK{ \setT}{Y} - \sum_{\setS\subseteq \setX} (-1)^{|\mathbf{S}|+1} \IcapVKK{ \setS}{Y} \; .
\end{eqnarray}

The terms that $\setS$ enumerates over is a subset of the terms that $\setT$ enumerates.  Therefore the $\sum_{\setS \subseteq \setX}$ completely cancels, leaving,

\begin{equation}
    \label{eq:dupsynergy}
    0 = \sum_{\setT \subseteq \setX} (-1)^{|\setT|} \IcapVKK{ \{X_1, T_1, \ldots, T_{|\setT|}\}}{Y} \; .
\end{equation}

If $\IcapVK$ obeys \textbf{(M)}, then each term of eq.~\eqref{eq:dupsynergy} s.t. $X_1 \not\in \setT$ cancels with the same term but with $X_1 \in \setT$.  This makes eq.~\eqref{eq:dupsynergy} sum to zero, and completes the proof.  Note we don't explicitly prove that $\IcapVK$ satisfies \textbf{(M)}, but if it does, then duplicate predictors do not increase synergy.

\subsection{Proof of bounds of $\Svkk{\setX}{Y}$}
\label{appendix:boundsproof}
We show that,
\begin{equation}
	\WMS\left( \setX : Y \right) \leq \Svk\left( \setX : Y \right) \leq \S_{\max}\left( \setX : Y \right) \; .
\end{equation}

\subsubsection{Proof that $\Svkk{ \setX }{Y} \leq \S_{\max}\left( \setX : Y \right)$}

We invoke the standard definitions of $\Svk$ and $\S_{\max}$,
\begin{eqnarray*}
	\Svkk{ \setX }{Y} &\equiv& \info{\X}{Y} - \Ivkk{ \setX }{Y} \\ 
	\S_{\max}( \setX : Y ) &\equiv& \info{\X}{Y} - \opI_{\max}( \setX : Y ) \; ,
\end{eqnarray*}

where $\Ivk$ and $\opI_{\max}$ are defined as,
\begin{eqnarray}
\nonumber	\Ivkk{\setX}{Y} &=& \mathbb{E}_Y \Ivkk{ \setX }{Y=y} \\
	&=& \mathbb{E}_Y \min_{\substack{p^*(X_1, \ldots, X_n, Y) \\ p^*(X_i,Y) = p(X_i,Y)  \quad \forall i}} \opI^*(\X : Y = y ) \\
	\opI_{\max}\left( \setX : Y \right) &\equiv& \mathbb{E}_Y \max_i \info{X_i}{Y=y} \; .
\end{eqnarray}

Now we prove $\Svkk{ \setX}{Y} \leq \S_{\max}( \setX : Y )$ by showing that $\Ivkk{\setX}{Y}~\geq~\opI_{\max}( \setX~:~Y )$.

\begin{proof}
\begin{eqnarray}
	\mathbb{E}_Y \Ivkk{\setX}{Y=y} &\geq& \mathbb{E}_Y \opI_{\max}\left( \setX : Y = y \right) \\
	\mathbb{E}_Y \left[ \Ivkk{\setX}{Y=y} - \opI_{\max}\left( \setX : Y = y \right) \right] &\geq& 0 \; .
\end{eqnarray}

Now expanding $\Ivkk{\setX}{Y=y}$ and $\opI_{\max}(\setX : Y = y)$,
\begin{equation}
\mathbb{E}_Y \left[ \left(\min_{\substack{p^*(X_1,\ldots,X_n,Y) \\ p^*(X_i,Y) = p(X_i,Y) \quad \forall i}} \opI^*(\X:Y = y) \right) - \max_i \info{X_i}{Y=y} \right] \geq 0 \; .
\end{equation}

We define the index $m \in \{1, \ldots, n\}$ such that $m~=~\argmax_i \info{X_i}{Y=y}$.  The predictor with the most information about state $Y=y$ is thus $X_m$.  This yields,

\begin{equation}
\mathbb{E}_Y \left[ \left(\min_{\substack{p^*(X_1,\ldots,X_n,Y) \\ p^*(X_i,Y) = p(X_i,Y) \quad \forall i}} \opI^*(\X:Y = y) \right) - \info{X_m}{Y=y} \right] \geq 0 \; .
\end{equation}

The constraint $p^*(X_i,Y) = p(X_i,Y)$ entails that $\info{X_m}{Y=y} = \opI^*(X_m~:~Y~=~y)$.  Therefore we can pull $\info{X_m}{Y=y}$ inside the minimization as a constant,
\begin{equation}
\mathbb{E}_Y \left[ \min_{\substack{p^*(X_1,\ldots,X_n,Y) \\ p^*(X_i,Y) = p(X_i,Y) \quad \forall i}} \opI^*(\X\!:\!Y~=~y) - \opI^*(X_m:Y=y) \right] \geq 0 \; .
\end{equation}
As $X_m$ is a subset of predictors $\X$, we can subtract it yielding,
\begin{equation}
\mathbb{E}_Y \left[ \min_{\substack{p^*(X_1,\ldots,X_n,Y) \\ p^*(X_i,Y) = p(X_i,Y) \quad \forall i}} \opI^*\left(X_{1\ldots n \setminus m}: Y = y \middle| X_m\right) \right] \geq 0 \; .
\end{equation}

The state-dependent conditional mutual information $\opI^*\!\left( X_{1 \ldots n \setminus m} : Y = y \middle| X_m \right)$ is a Kullback-Liebler divergence.  As such it is nonnegative.  Likewise the minimum of a nonnegative quantity is also nonnegative.

\begin{equation}
\mathbb{E}_Y \left[ \underbrace{ \min_{\substack{p^*(X_1,\ldots,X_n,Y) \\ p^*(X_i,Y) = p(X_i,Y) \quad \forall i}} \opI^*\left(X_{1\ldots n \setminus m}: Y = y \middle| X_m\right) }_{\geq 0} \right] \geq 0 \; .
\end{equation}

Finally, the expected value of a list of nonnegative quantities is nonnegative.  And the proof that $\Svkk{ \setX }{Y} \leq \S_{\max}( \setX : Y )$ is complete.

\end{proof}

\clearpage
\subsubsection{Proof that $\WMS( \setX : Y ) \leq \Svkk{ \setX }{Y}$}
We invoke the standard definitions of $\WMS$ and $\Svk$,
\begin{eqnarray}
	\WMS( \setX : Y ) &\equiv& \info{\X}{Y} - \sum_{i=1}^n \info{X_i}{Y} \\
	\Svkk{ \setX }{Y} &\equiv& \info{\X}{Y} - \Ivkk{ \X }{Y} \\ 
	&=& \info{\X}{Y} - \VKbox \opI^*( \X : Y ) \; .
\end{eqnarray}

We prove the conjecture $\WMS( \setX : Y ) \leq \Svkk{ \setX }{Y}$ by showing,
\begin{equation}
	\VKbox \opI^*(\X:Y) \leq \sum_{i=1}^n \info{X_i}{Y} \; .
\end{equation}

Given:
\begin{equation}
	\label{eq:wmsbound1}
	\min_{\substack{p^*(X_1,\ldots,X_n,Y) \\ p^*(X_1,Y) = p(X_1,Y) \\ \vdots \\ p^*(X_n,Y) = p(X_n,Y) }} \opI^*(\X:Y) \; ,
\end{equation}

the individual constraint $p^*(X_1,Y) = p(X_1,Y)$ can add at most $\info{X_1}{Y}$ bits to $\opI^*\left(\X : Y \right)$.  Therefore we can upperbound eq.~\eqref{eq:wmsbound1} by dropping the constraint $p^*(X_1,Y) = p(X_1,Y)$ and adding $\info{X_1}{Y}$.  This yields,

\begin{equation}
	\label{eq:wmsbound2}
	\IcupVKK{ \setX }{Y} \leq \min_{\substack{p^*(X_1,\ldots,X_n,Y) \\ p^*(X_2,Y) = p(X_2,Y) \\ \vdots \\ p^*(X_n,Y) = p(X_n,Y) }} \opI^*(\X:Y) + \info{X_1}{Y} \; .
\end{equation}

Likewise, the righthand-side of eq.~\eqref{eq:wmsbound2} can be upperbounded by dropping the constraint $p^*(X_2,Y) = p(X_2,Y)$ and adding $\info{X_2}{Y}$.  This yields,

\begin{equation}
	\label{eq:wmsbound3}
\min_{\substack{p^*(X_2,\ldots,X_n,Y) \\ p^*(X_2,Y) = p(X_2,Y) \\ \vdots \\ p^*(X_n,Y) = p(X_n,Y) }} \opI^*(\X:Y) \leq \min_{\substack{p^*(X_3,\ldots,X_n,Y) \\ p^*(X_3,Y) = p(X_3,Y) \\ \vdots \\ p^*(X_n,Y) = p(X_n,Y) }} \opI^*(\X:Y) + \info{X_1}{Y} + \info{X_2}{Y} \; .
\end{equation}

Repeating this process $n$ times yields,

\begin{eqnarray}
	\label{eq:wmsbound4}
	\Ivkk{ \setX }{Y} &\leq& \min_{p^*(X_1, \ldots, X_n,Y) } \opI^*( \X : Y )  + \sum_{i=1}^n \info{X_i}{Y} \\
	&=& \sum_{i=1}^n \info{X_i}{Y} \; .
\end{eqnarray}

\clearpage
\section{Algebraic simplification of $\Delta I$}
\label{appendix:deltai}

Prior literature \cite{nirenberg01, nirenberg03, pola03, latham-05} defines $\Delta \opname{I}\left( \setX ; Y \right)$ as,

\begin{eqnarray}
\label{eq:deltai}
\Delta \opname{I}\left( \setX ; Y \right) &\equiv& \DKL{\Prob{Y|\X}} { {\textstyle \Pr_{\ind}}\left(Y \middle|\setX\right)  } \\
&=& \label{eq:average_deltai} \sum_{\setx, y \in \setX, Y} \Prob{\setx, y} \log \frac{\Prob{y|\setx}}{\Pr_{\ind}(y|\setx) } \; .
\end{eqnarray}

Where,
\begin{eqnarray}
	\textstyle{\Pr_{\ind}}(Y = y| \setX = \setx) &\equiv& \frac{\Prob{y} \Pr_{\ind}(\setX = \setx| Y = y) }{\Pr_{\ind}(\setX = \setx)} \\
	&=& \displaystyle \frac{\Prob{y} \prod_{i=1}^n \Prob{ x_i | y} }{ \Pr_{\ind}(\setx) } \\
	\textstyle{\Pr_{\ind}}(\setX = \setx) &\equiv& \sum_{y \in Y} \Prob{Y = y} \prod_{i=1}^n \Prob{x_i|y}
\end{eqnarray}

The definition of $\Delta \opname{I}$, eq.~\eqref{eq:deltai}, reduces to,

\begin{eqnarray}
	\Delta \opname{I}\left( \setX ; Y \right) &=& \sum_{\setx, y \in \setX, Y} \Prob{\setx, y} \log \frac{\Prob{y|\setx}}{\Pr_{\ind}(y|\setx) } 	\\
	&=& \sum_{\setx, y \in \setX, Y} \Prob{\setx, y} \log \frac{\Prob{y|\setx} \Pr_{\ind}(\setx) }{ \Prob{y} \prod_{i=1}^n \Prob{x_i|y} } \\
	&=& \sum_{\setx, y \in \setX, Y} \Prob{\setx, y} \log \frac{\Prob{\setx|y}}{\prod_{i=1}^n \Prob{x_i|y} }	\frac{\Pr_{\ind}(\setx)}{\Prob{\setx}} \\
\nonumber	&=& \sum_{\setx, y \in \setX, Y} \Prob{\setx, y} \log \frac{\Prob{\setx|y}}{\prod_{i=1}^n \Prob{x_i|y} }	+ \sum_{\setx, y \in \setX, Y} \Prob{\setx, y} \log \frac{\Pr_{\ind}(\setx)}{\Prob{\setx}} \\
	&=& \sum_{\setx, y \in \setX, Y} \Prob{\setx, y} \log \frac{\Prob{\setx|y}}{\prod_{i=1}^n \Prob{x_i|y} }	- \sum_{\setx \in \setX} \Prob{\setx} \log \frac{\Prob{\setx}}{\Pr_{\ind}(\setx)} \\
\nonumber	&=& \DKL{ \Prob{\X|Y} }{ \prod_{i=1}^n \Prob{X_i|Y}} - \DKL{ \Prob{\X} }{ {\textstyle \Pr_{\ind}}( \setX ) } \\
	&=& \opname{TC}\left( X_1; \cdots ; X_n \middle| Y \right) - \DKL{ \Prob{\X} }{ {\textstyle \Pr_{\ind}}( \setX ) } \; .
\end{eqnarray}

where $\opname{TC}\left( X_1; \cdots ; X_n \middle| Y \right)$ is the conditional total correlation among the predictors given $Y$.

\end{document}

%% file: this_symbols.tex
\usepackage{nips10submit_e}
\nipsfinalcopy

\definecolor{mygray}{RGB}{153,153,153}
\definecolor{myred}{RGB}{221,77,57}
\definecolor{myblue}{RGB}{68,104,252}
\definecolor{mygreen}{RGB}{86,149,22}

\newcommand*{\X}{\ensuremath{X_{1 \ldots n}\xspace}}

\newcommand*{\bin}[1]{\texttt{#1}}

\newcommand*{\WMS}{\operatorname{WMS}}

\renewcommand*{\S}{\ensuremath{\mathcal{S}}\xspace}

\newcommand*{\setX}{\ensuremath{\mathbf{X}}\xspace}
\newcommand*{\setS}{\ensuremath{\mathbf{S}}\xspace}
\newcommand*{\setT}{\ensuremath{\mathbf{T}}\xspace}
\newcommand*{\setx}{\ensuremath{\mathbf{x}}\xspace}

\newcommand*{\Probstar}[1]{ \ensuremath{ \textstyle{\Pr^*} \! \left( #1 \right)}}

\newcommand*{\infostar}[2]{ \operatorname{I}^* \! \left( #1 \!:\! #2 \right)}

\newcommand*{\Ivk}{\operatorname{I}_{\textsc{VK}}\xspace}
\newcommand*{\Ivkk}[2]{\operatorname{I}_{\textsc{VK}}\!\left( #1 ~:~ #2 \right)}

\newcommand*{\IcapVK}{\ensuremath{\opname{I}_{\cap}^{\textsc{VK}} \xspace}}
\newcommand*{\IcapVKK}[2]{\ensuremath{\opname{I}_{\cap}^{\textsc{VK}} \xspace}\!\left( #1 \! : \! #2 \right)}

\newcommand*{\IcupVK}{\ensuremath{\opname{I}_{\textsc{VK}} \xspace}}
\newcommand*{\IcupVKK}[2]{\ensuremath{\opname{I}_{\textsc{VK}} \xspace}\!\left( #1 \! : \! #2 \right)}

\newcommand*{\Svk}{\mathcal{S}_{\textsc{VK}}}
\newcommand*{\Svkk}[2]{{\mathcal{S}_{\textsc{VK}}\!\left( #1 \!:\! #2 \right)}}

\newcommand*{\VKbox}{ \min_{\substack{p^*(X_1, \ldots, X_n,Y) \\ p^*(X_i,Y) = p(X_i,Y) \quad \forall i}} }

%% file: this_authors.tex
\usepackage{authblk}
\author[1,*]{Virgil Griffith}
\author[1,2]{Christof Koch}
\affil[1]{Computation and Neural Systems, Caltech, Pasadena, CA 91125}
\affil[2]{Allen Institute for Brain Science, Seattle, WA 98103}

\let\oldthefootnote\thefootnote
\renewcommand{\thefootnote}{\fnsymbol{footnote}}
\footnotetext[1]{To whom correspondence should be addressed. Email: {\tt virgil@caltech.edu}}
\let\thefootnote\oldthefootnote